\documentclass[aps,prb,showpacs, twocolumn, 10pt]{revtex4-1}
\pdfoutput=1
\usepackage{times}
\usepackage{epsfig}
\usepackage{amsmath}
\usepackage{amssymb}
\usepackage{wasysym}
\usepackage{MnSymbol}
\usepackage[usenames,dvipsnames]{xcolor}
\usepackage[colorlinks=true,citecolor=blue,linkcolor=blue]{hyperref}
\newcommand{\bs}{\boldsymbol}

\begin{document}

\title{Electronic structure of (LaNiO$_3$)$_2$/(LaAlO$_3$)$_N$ heterostructures grown along [111]}
\author{Andreas R\"uegg}
\author{Chandrima Mitra}
\author{Alexander A. Demkov}
\author{Gregory A. Fiete}
\affiliation{Department of Physics, The University of Texas at Austin, Austin, Texas 78712, USA}
\begin{abstract}
The electronic structure of a LaNiO$_3$ bilayer grown along the [111] direction and confined between insulating layers of LaAlO$_3$ is theoretically investigated using a combination of first principle calculations and effective multi-orbital lattice models. The LDA band structure is well reproduced by a tight-binding model for the Ni-$e_g$ orbitals defined on the buckled honeycomb lattice. We highlight peculiar properties of this model which include almost flat bands as well as linear and quadratic band crossing points. The effect of local correlations is discussed within the LDA$+U$ scheme and within the Hartree-Fock approximation for interacting multi-orbital lattice models. Over a wide range of interaction parameters we find that a ferromagnetic phase is energetically favored. We discuss the possibility of additional orbital order which could stabilize a spontaneous Chern insulator with chiral edge modes or a staggered orbital phase with a $\sqrt{3}\times\sqrt{3}$ reconstruction of the unit cell. By studying an interacting nickel-oxygen lattice model we find that the stability of these orbitally ordered phases also depends on the value of the charge-transfer energy. Controlling the charge-transfer energy might therefore be an important step towards engineering exotic electronic phases in certain classes of oxide heterostructures.
\end{abstract}
\date{\today}
\maketitle
\section{Introduction}
Design, growth and characterization of artificial structures of complex oxide materials has become an important field of research in modern condensed matter physics.\cite{Zubko:2011,Chakhalian:2012} The wide range of available materials with a multitude of physical properties carry great potential to engineer novel electronic devices\cite{Mannhart:2010} with desired functionalities, such as superconductivity, ferromagnetism, ferroelectricity or multiferroic behavior. There are various physical mechanisms which modify the electronic properties at interfaces and in superlattices as opposed to the bulk systems which can lead to atomic, electronic\cite{Ohtomo:2002,Okamoto:2004a} and orbital reconstructions.\cite{Chakhalian:2007}

A promising strategy to tune the electronic properties at interfaces and in superlattices of complex oxides is to grow the heterostructure along uncommon crystallographic directions, such as along the [111] direction of a cubic perovskite.\cite{Xiao:2011} One interesting feature of such structures is the possibility to create a large variety of effectively two-dimensional lattices by sandwiching a precise number of layers of an ``active material" between a good insulator. For example, a (111) bilayer of a cubic lattice forms a (buckled) honeycomb lattice while three adjacent layers form a three-dimensional version of the dice lattice.\cite{Xiao:2011,Yang:2011,WangFa:2011,Ruegg:2011c} 
Realizing these lattice geometries is particularly interesting in view of engineering electronic phases with topologically non-trivial band structures such as Chern insulators\cite{Haldane:1988} or time-reversal invariant topological insulators\cite{Kane:2005b} with their associated protected surface modes.\cite{Xiao:2011,Ruegg:2011c,Yang:2011}
\begin{figure}
\includegraphics[width=1\linewidth]{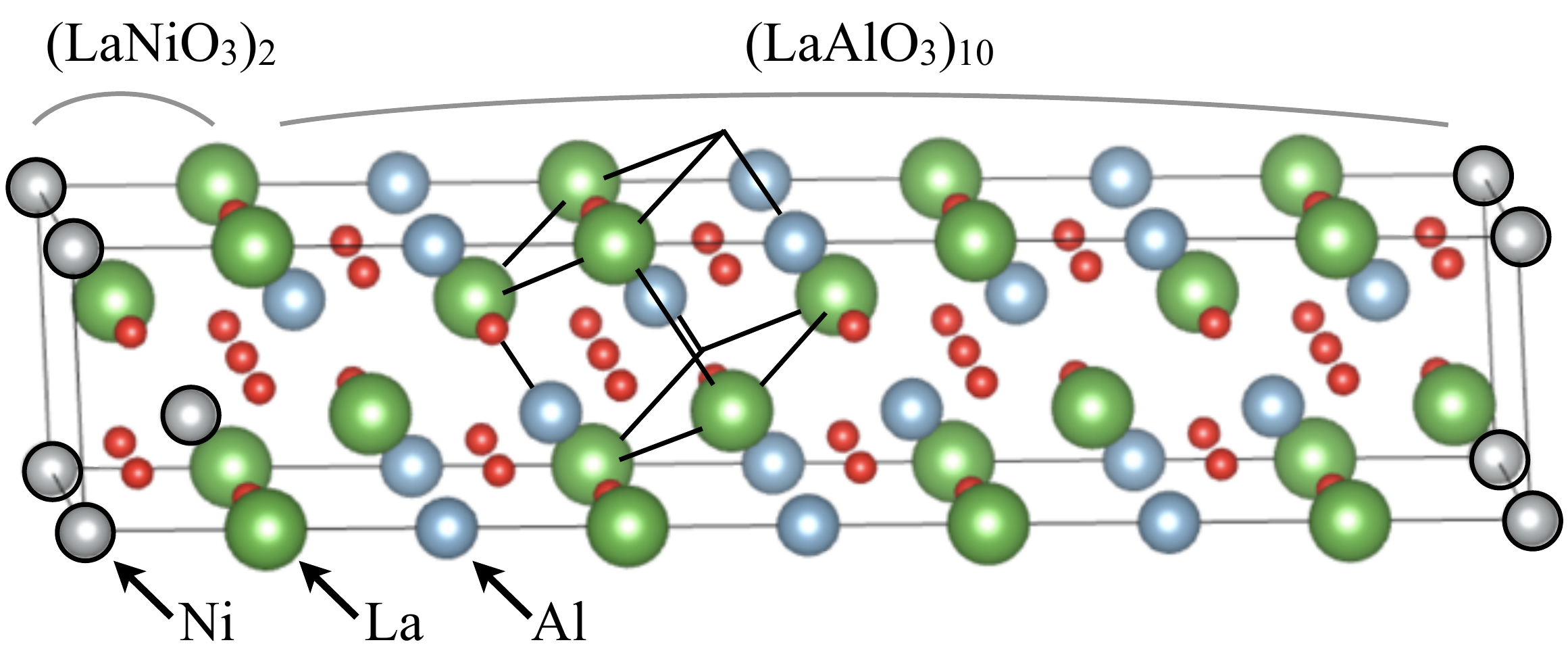}
\caption{The supercell of the superlattice (LaNiO$_3$)$_2$/(LaAlO$_3$)$_{10}$ considered in this work. The growth direction is the [111] direction of the cubic perovskite. The primitive cell of the perovskite structure is also indicated.}
\label{fig:SL}
\end{figure}

In this article, we discuss the electronic structure of a (111) bilayer of LaNiO$_3$ sandwiched between sufficiently large regions of LaAlO$_3$. LaAlO$_3$ is a wide band-gap insulator\cite{Seo:2011} with an experimentally determined gap of $E_g = 5.6$~eV while LaNiO$_3$ is a paramagnetic metal in bulk. The formal valence of nickel is Ni$^{3+}$ which corresponds to a partially filled $3d$-shell with an electronic configuration $t_{2g}^6e_g^1$, i.e. one electron in the doubly degenerate $e_g$ manifold. The insulating LaAlO$_3$ layers insure that the electronic degrees near the Fermi energy are effectively confined to the buckled honeycomb lattice formed by the Ni ions of the (111) bilayer. Figure~\ref{fig:SL} shows the actual supercell (LaNiO$_3$)$_2$/(LaAlO$_3$)$_{10}$ used in our density functional theory (DFT) calculations.\cite{Hohenberg:1964,Kohn:1965}

Digital heterostructures between LaNiO$_3$ and LaAlO$_3$ have been considered previously but the focus was on systems grown along the [001] direction.\cite{Liu:2011,Boris:2011,Freeland:2011,Chakhalian:2011} The interest in the [001]-grown heterostructures has been stimulated (at least in part) by the prospect to control the orbital character of the conduction electrons by applying tensile or compressive strain. As opposed to bulk LaNiO$_3$ where no sign of orbital preference is observed, the reduced symmetry in the [001] heterostructure splits the energy between the $d_{z^2}$ and the $d_{x^2-y^2}$ orbitals. Theoretically, it has been suggested to apply tensile strain to selectively lower the energy of the $d_{x^2-y^2}$-orbital. In this case, one could hope to imitate the relevant conditions for cuprate high-temperature superconductivity with nickelate analogs.\cite{Anisimov:1999,Chaloupka:2008,Hansmann:2009} However, instead of a phase with uniformly occupied $d_{x^2-y^2}$ orbitals, a charge-density wave is experimentally observed for tensile strain,\cite{Liu:2011,Boris:2011,Freeland:2011} similar to the insulating bulk phases observed in other nickelates RNiO$_3$ with R=Pr, Nd, or Sm.\cite{Zhou:2003}

In contrast to the [001] heterostructure, the $e_g$-orbitals of the Ni ions remain doubly degenerate for the heterostructure grown along the [111] direction as long as the trigonal symmetry is preserved. The non-interacting band structure of an $e_g$-orbital model has a degeneracy point at ${\bs k}=0$ where two bands touch quadratically. Such a situations can in principle give rise to interesting weak-coupling instabilities, including spin nematic and topological insulator phases.\cite{Sun:2009,Ruegg:2011c} From the theoretical point of view, strong electronic correlations, charge-transfer physics and the complex interplay between the electronic and lattice degrees of freedom pose many challenges in modeling the nickelates. Moreover, to the best of our knowledge, there are currently no experimental data available for the (111) bilayer system. In this study, we use natural idealizations of the real system which allow us to understand the electronic structure from simple multi-orbital lattice models. However, due to the lack of phenomenological input, the present work should be considered as a first step towards a more accurate theoretical description of these interesting structures.

This paper is organized as follows. In Sec.~\ref{sec:DFT} we present DFT results of the electronic structure for the supercell in Fig.~\ref{fig:SL}. We show that the electronic degrees near the Fermi energy are predominately given by the $e_g$ electrons with mixing of the nearby oxygen-$p$ electrons. Effects of local correlations are discussed within the LSDA$+U$ scheme\cite{Dudarev:1998} which predicts ferromagnetism. In Sec.~\ref{sec:eg_model} we show that the band structure is well reproduced by a tight-binding model for $e_g$ electrons on the buckled honeycomb lattice and highlight universal properties arising from the particular geometry. The effect of local correlations is considered within this reduced model using a conventional mean-field approximation. In agreement with the LSDA$+U$ calculation, we find that a ferromagnetic phase is stabilized over a wide range of interaction parameters. Depending on the relative strength of the Hund's coupling $J$, we also identify spontaneous orbital order. In Sec.~\ref{sec:CT} we introduce an effective model which includes both the Ni-$e_g$ and oxygen-$p$ states. This allows us to study how the charge-transfer physics\cite{Zaanen:1985} alters the stability of the orbitally ordered phases.

\section{Density functional theory}
\label{sec:DFT}
We have studied the electronic structure of the (LaNiO$_3$)$_2$/(LaAlO$_3$)$_{10}$ supercell (see Fig.~\ref{fig:SL}) using density functional theory (DFT)\cite{Hohenberg:1964,Kohn:1965} within the local density approximation (LDA)\cite{Kohn:1965} as well as the local spin density approximation (LSDA)\cite{Barth:1972} as implemented in the Vienna ab-initio simulation package (VASP).\cite{Kresse:1996} We used the projector augmented wave pseudopotentials for all our calculations.\cite{Blochl:1994} A plane wave cut off energy of 600 eV and a $6\times6\times6$ ${\bs k}$-point grid was chosen for integrating over the Brillouin zone. The lattice constant for the supercell was chosen as {3.82 \AA}  which corresponds to the experimental pseudocubic lattice constant of bulk LaAlO$_3$. Atomic relaxation effects were not taken into account for our calculations.

\begin{figure}
\includegraphics[width=0.7\linewidth]{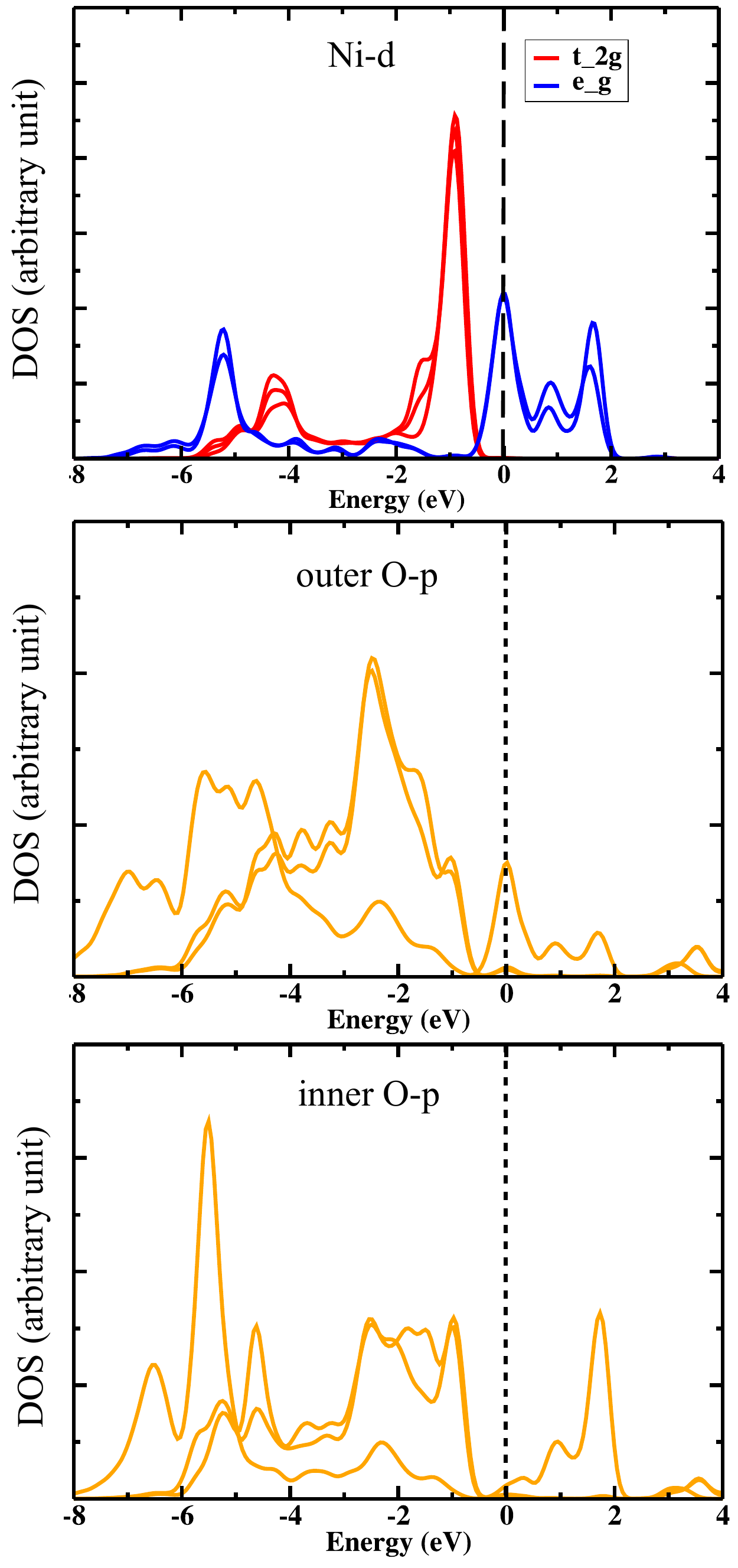}
\caption{(Color online.) LDA result for the orbitally projected  of the Ni ions and its two neighboring O ions which are distinguished by symmetry.}
\label{fig:LDA_DOS}
\end{figure}

\begin{figure}
\includegraphics[width=1\linewidth]{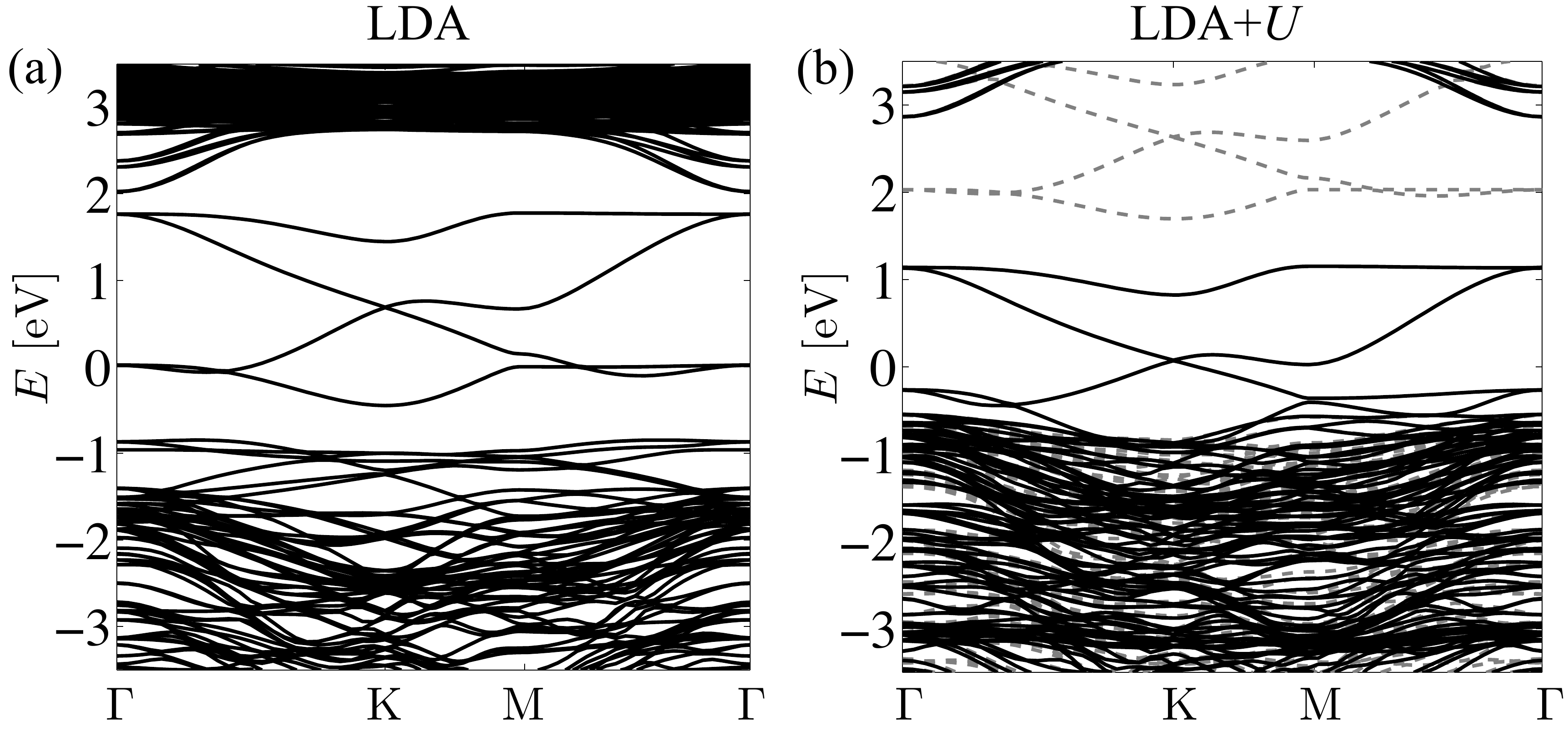}
\caption{(a) The LDA band structure near the Fermi level along a high-symmetry path in the hexagonal Brillouin zone. (b) The LDA+$U$ band structure along the same path as in (a). The solid (dashed) curves denote the energy bands of the majority (minority) spin projection. The zero of energy is set to the Fermi level.}
\label{fig:bandstructures}
\end{figure}

The LDA band gap for LaAlO$_3$ is $E_g=3.3$~eV (experimental value $E_g=5.6$~eV).\cite{Seo:2011} As expected, this wide band gap leads to a strong confinement of the electronic degrees of freedom to the LaNiO$_3$ bilayer. The spectral weight near the Fermi energy is predominately comprised of states located at the Ni and their neighboring O ions. On the adjacent Al ion, we find very little spectral weight near the Fermi energy. These states on the Al are mainly associated with the $s$ and $p$ orbitals which weakly hybridize with the O-$p$ states connecting the nearby Ni ion. We do not find a recognizable difference in the total charge of an Al ion located next to the interface or far away, indicating that there is essentially no charge transfer from the Ni to the Al. By symmetry, the ``inner O layer" which is sandwiched between the two Ni layers is distinguished from the two ``outer O layers" adjacent to only one of the two Ni layers. Figure~\ref{fig:LDA_DOS} shows the orbital projected  for a Ni atom as well as for an inner and outer O as obtained within the LDA. The LSDA calculations gave identical results. In Fig.~\ref{fig:LDA_DOS} we set the zero of the energy to the Fermi level. The octahedral crystal field splits the Ni 3$d$ orbitals into the triply degenerate $t_{2g}$ states and the doubly degenerate $e_g$ states which are well separated near the Fermi level. The $e_g$ DOS near the Fermi energy has a three peak structure with two prominent features near $E\approx0$ and $E\lesssim2$ eV. As we show in Sec.~\ref{sec:eg_model}, these peaks appear due to weakly dispersing bands which are generically present in an $e_g$ model on the buckled honeycomb lattice. The orbital projected  for the O atoms show that there is considerable mixing between the Ni-$e_g$ and the O-$p$ states. Out of the three $p$ orbitals it is the $p_{\sigma}$ orbital pointing towards the Ni ion which hybridizes most strongly. Also note that the outer O mixes strongly with the lower and the inner O with the upper flat band.
The LDA band structure along a high-symmetry path in the Brillouin zone is shown in Fig.~\ref{fig:bandstructures}(a). Near the Fermi level, there is a well-separated block consisting of four bands for each spin projection. Comparing with the results for the projected DOS in Fig.~\ref{fig:LDA_DOS} we can associate these four bands with the two $e_g$ orbitals and the A-B sublattice structure of the (111) bilayer. We will discuss further characteristic properties of this band structure in more detail in Sec.~\ref{sec:eg_model}.

In order to study correlation effects of the $d$ electrons in Ni we also performed LSDA$+U$ calculations where we employed the simplified rotationally invariant approach as introduced by Dudarev et al.\cite{Dudarev:1998} We choose an effective $U_{\rm eff}$ value of 5.74 eV which was taken from Ref.~\onlinecite{Gou:2011} where the $U_{\rm eff}$ parameter has been computed in a self-consistent approach following the scheme introduced in Ref.~\onlinecite{Cococcioni:2005}. It has been shown that this choice leads to accurate results for the ground state atomic structure.\cite{Gou:2011} The chosen $U_{\rm eff}$ is also consistent with the reported experimental value in a recent work by Nohara {\it et.~al}\cite{Nohara:2009} who performed GW as well as LSDA+$U$ calculations on LaNiO$_3$. Upon fitting the calculated energy spectra of a cluster model to experimental X-ray photoelectron spectroscopy (XPS) and X-ray absorption spectroscopy (XAS) data\cite{Abbate:2002} an effective $U_{\rm eff}$ of 5.7 eV was obtained.

\begin{figure}
\includegraphics[width=0.7\linewidth]{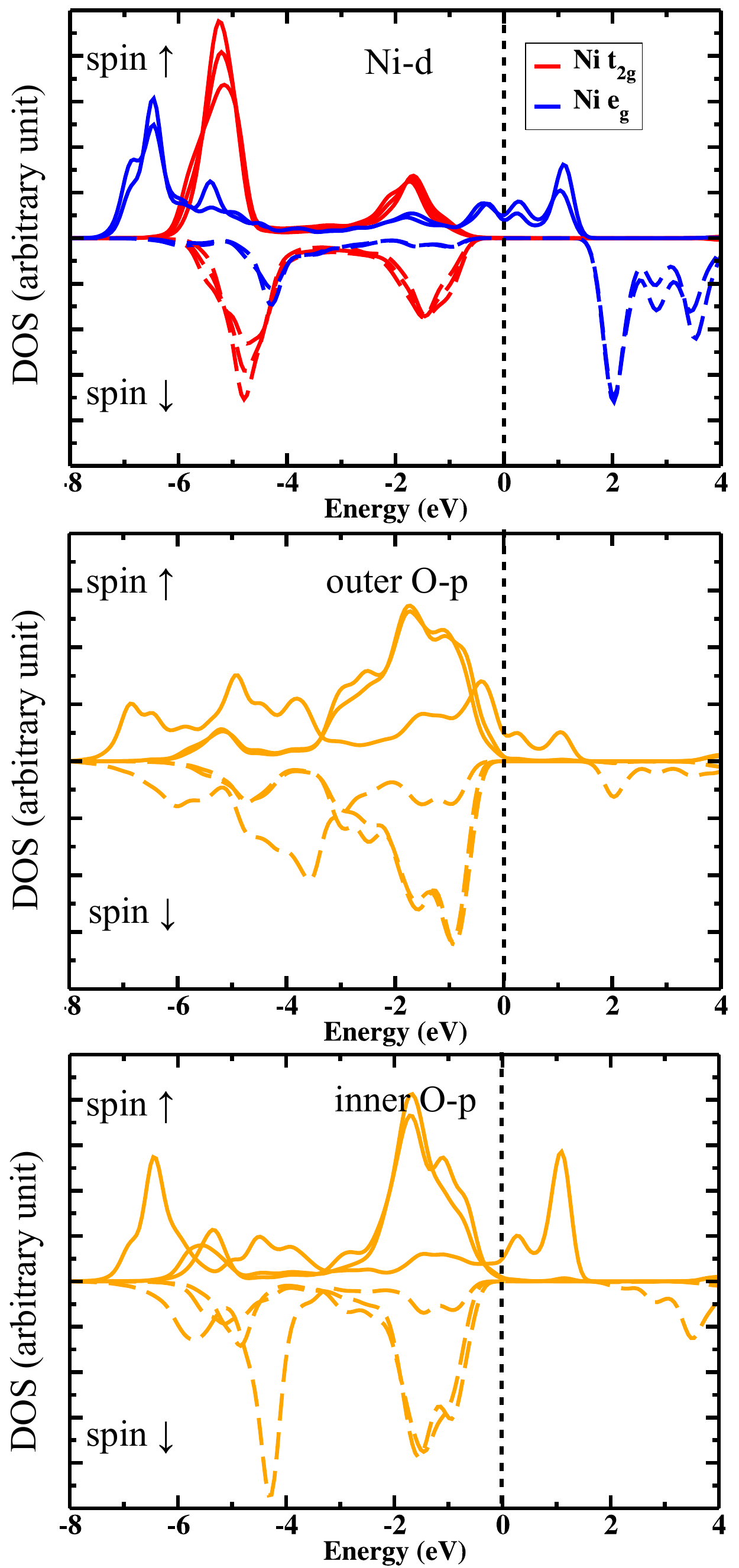}
\caption{(Color online.) LDA$+U$ results for the orbitally projected  in the ferromagnetic phase. Shown are the Ni 3$d$ orbitals and the $p$ orbitals for the two neighboring O ions which are distinguished by symmetry. Spin-$\uparrow$ and spin-$\downarrow$ contributions are shown separately.}
\label{fig:LDAU_DOS}
\end{figure}

Our LSDA+$U$ calculation favors a ferromagnetic state with a magnetic moment of $1.12\mu_{\rm B}$ on each Ni atom, where $\mu_{\rm B}$ is the Bohr magneton. This is consistent with the theoretical value of $1\mu_{\rm B}$ from LSDA+$U$ reported in Ref.~\onlinecite{Gou:2011} for bulk LaNiO$_3$. Figure~\ref{fig:bandstructures}(b) shows the spin-resolved  band structure near the Fermi level. The $e_g$ bands for spin-$\uparrow$ and spin-$\downarrow$ completely separate. The Fermi level is in the fully polarized majority band and is now shifted to the vicinity of the linear band-crossing point (Dirac point) located at K (and K'). Figure~\ref{fig:LDAU_DOS} shows the projected and spin-resolved DOS. Note that the ferromagnetic solution remains gapless and is fully polarized such that only the majority electrons contribute to the DOS at the Fermi energy. In our LSDA$+U$ calculation there is no sign of orbital ordering. 

Bulk LaNiO$_3$ is experimentally found to be paramagnetic although LSDA$+U$ calculations predict a ferromagnetic phase.\cite{Gou:2011} For the bulk system, it has been argued that the poor treatment of the dynamical screening in the LSDA+$U$ is responsible for the disagreement with experiments.\cite{Gou:2011,Nohara:2009} From a phenomenological point of view, $U_{\rm eff}=0$ gives the best agreement with experimental data\cite{Gou:2011} and this choice has been used as starting point for GW calculations.\cite{Nohara:2009}
The ferromagnetic state found in our LSDA+$U$ calculation (with $U_{\rm eff}=5.74$eV) for the bilayer system should therefore be taken with care. On the one hand, one expects that the static treatment of the atomic interaction could be even more problematic in the quasi-two dimensional bilayer system. On the other hand, ordered phases not present in the bulk have been experimentally observed in strained LaNiO$_3$ thin films.\cite{Chakhalian:2011} Due to the lack of phenomenological input from experiments, we think that this uncertainty can not be resolved at the present moment. 
\section{$e_g$-orbital model for the (111) bilayer}
\label{sec:eg_model}
Let us now discuss the LDA band structure from the perspective of a tight-binding model defined on the buckled honeycomb lattice. The tight-binding model helps to identify generic features of the band structure in this particular geometry. Based on the observations made in the previous section, we focus here on an orbital model for $e_g$ electrons alone. This is justified if the crystal field splitting is sufficiently large and the energy of the oxygen $p$-states is well separated from the energy of the Ni-$e_g$ states. As we will show in the following, this approach allows us to reproduce the LDA band structure fairly well. Nevertheless, the $e_g$ model discussed in this section misses an important physical ingredient, namely the charge transfer between the Ni and O ions. In Sec.~\ref{sec:CT} we therefore review the role of the charge transfer in the context of a more general model which also includes oxygen $p$ states.

\subsection{Idealized tight-binding model}
\begin{figure}
\includegraphics[width=0.9\linewidth]{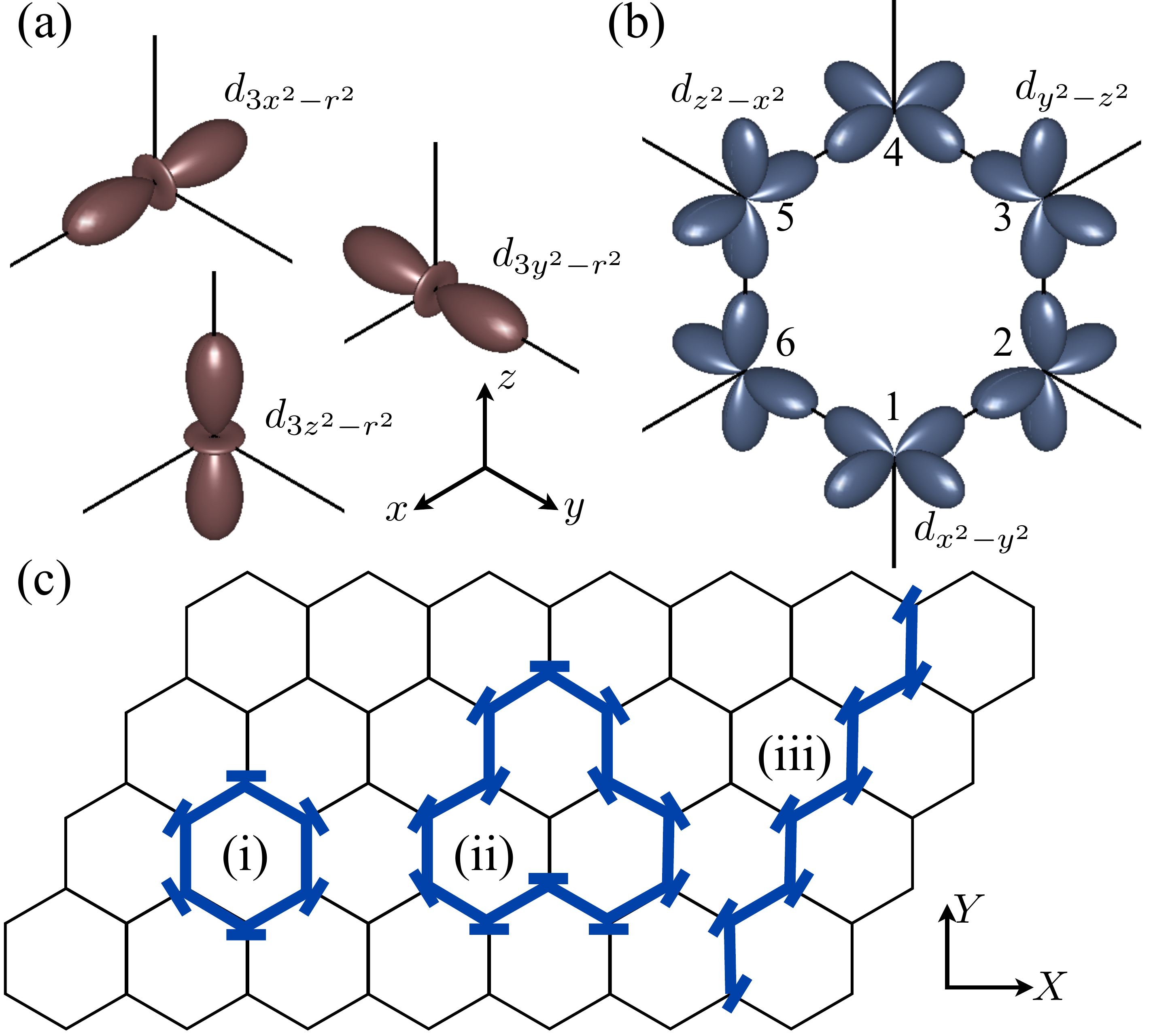}
\caption{(a) Oxygen mediated hopping along the principle axis of the cube are between the $d_{3u^2-r^2}$ orbitals with $u=x,y,z$. (b) A spatially localized eigenstate of the nearest-neighbor tight-binding model Eq.~\eqref{eq:Ht}. The viewpoint is along [111]. (c) Taking equal weight superpositions of the hexagonal states (i) allows one to construct localized states with support on arbitrary contractible loops (ii). There are two more states which are linearly independent from the hexagonal states and have support on a loop encircling the torus in one or the other direction (iii).}
\label{fig:tb_orbitals}
\end{figure}

In the simplest tight-binding model we consider only the dominant hopping between neighboring transition-metal ions which is mediated via an intermediate oxygen $p$-orbital. It is convenient to define the following orbitals
\begin{equation}
d_{u^2}\equiv d_{3u^2-r^2}=\hat{e}_{u}\cdot\vec{d},\quad \vec{d}=(d_{x^2-y^2},d_{3z^2-r^2}),
\label{eq:du}
\end{equation}
where $\hat{e}_x=(\sqrt{3}/2,-1/2)$, $\hat{e}_y=(-\sqrt{3}/2,-1/2,)$ and $\hat{e}_z=(0,1)$ are the unit vectors in the directions of the nearest-neighbor bonds projected to the plane perpendicular to the [111] direction. The real orbitals $d_{x^2}$ and $d_{y^2}$ are rotations of the $d_{z^2}$ orbital around the [111] axis by $\pm 2\pi/3$ and are shown in Fig.~\ref{fig:tb_orbitals}(a). 
For oxygen mediated hopping in the direction $\hat{e}_u$, the Slater-Koster energy integrals\cite{Slater:1954} are only finite between neighboring $d_{u^2}$-orbitals. The tight-binding model takes the simple form $H_t=\sum_{\sigma}H_{t\sigma}$ where 
\begin{equation}
H_{t\sigma}=-t\sum_{i\in A}\sum_{u=x,y,z}\left(d_{i,u^2,\sigma}^{\dag}d_{i+\hat{e}_u,u^2,\sigma}^{}+\rm{h.c.}\right).
\label{eq:Ht}
\end{equation}
The operators $d_{i,u^2,\sigma}^{\dag}$ creates an electron with spin $\sigma$ in a $d_{u^2}$-orbital at site $i$ and the summation runs over the sites of the $A$ sublattice of the honeycomb lattice. Employing periodic boundary conditions, Eq.~\eqref{eq:Ht} is readily diagonalized in momentum space.\cite{Xiao:2011} The band structure shows two dispersing bands which are reminiscent of the electronic structure of graphene,
\begin{equation}
\varepsilon_2({\bs k})=-t/2\sqrt{3+A_{\bs k}},\quad \varepsilon_3({\bs k})=t/2\sqrt{3+A_{\bs k}}.
\end{equation}
Here, we have introduced
\begin{equation}
A_{\bs k}=2\cos(\sqrt{3}k_X)+4\cos(\frac{\sqrt{3}}{2}k_X)\cos(\frac{3k_Y}{2}).
\end{equation}
The subscript of the wave-vectors refers to the two-dimensional $(X,Y)$-coordinate system, see Fig.~\ref{fig:tb_orbitals}, and the unit of length is chosen as the length of the projection of the nearest-neighbor bond into the
[111] plane. The two dispersing bands cross linearly at ${\bs K}_+=\left(\frac{4\pi}{3\sqrt{3}},0\right)$ and ${\bs K}_-=-{\bs K}_+$ forming Dirac cones as in graphene. However, different from graphene, the dispersing bands are ``sandwiched" by two flat bands with energy
\begin{equation}
\varepsilon_1({\bs k})=-3t/2,\quad \varepsilon_4({\bs k})=3t/2.
\end{equation}
The two peaks observed in the projected  of the Ni $e_g$ orbitals in the LDA calculation, Fig.~\ref{fig:LDA_DOS}, can be related to these two flat bands.
The flat bands touch the dispersing bands in a single point at ${\bs k}=0$. Remarkably, the model Eq.~\eqref{eq:Ht} is formally equivalent to the planar $p$-band model on the honeycomb lattice introduced in Ref.~\onlinecite{Wu:2007}.\footnote{We would like to thank Adrian Feiguin for pointing out a possible connection.} However, the physical implementation of Eq.~\eqref{eq:Ht} is very different in the present context: instead of $p$-orbitals, the elementary degrees of freedom are the $e_g$-orbitals of the $d$-electrons which hop on a {\it buckled} honeycomb lattice formed by a cubic (111) bilayer instead of the {\it planar} hexagonal lattice. 

We can use this formal analogy to the planar $p$-model to give a physical interpretation of the flat bands and the degeneracy points. This is done by constructing spatially localized eigenstates of Eq.~\eqref{eq:Ht}.\cite{Wu:2007,Bergman:2008} Owing to the fact that hopping in the $\hat{e}_u$ direction is only finite between $d_{u^2}$-orbitals, it is possible to find spatially localized eigenfunctions which extend around a single hexagon located at ${\bs R}$, see Fig.~\ref{fig:tb_orbitals}(b):
\begin{equation}
|\Psi_{\bs R}^{\nu}\rangle=\frac{1}{\sqrt{6}}\sum_{j=1}^6(-\nu)^{j}\left(\cos\theta_j|d_{j,x^2-y^2}\rangle-\sin\theta_j|d_{j,z^2}\rangle\right).
\label{eq:PsiR}
\end{equation}
Here, $\nu=\pm$ labels a member of the bottom (-) and top (+) flat band manifold and $\theta_j=(j-1)\pi/3$. At each site $j$ of the hexagon, the orbital wave function is orthogonal to the ``outward-pointing" orbital $d_{j,u_j^2}$, which confines the electron to the hexagon. Assuming $N$ unit cells, there are a total of $N$ hexagon states for a given $\nu$. Taking equal weight superpositions of different hexagonal states, one can construct eigenstates of Eq.~\eqref{eq:Ht} with support on arbitrary contractible loops, see Fig.~\ref{fig:tb_orbitals}(c). 

From the definition of the hexagonal states Eq.~\eqref{eq:PsiR} it follows that $\sum_{\bs R}|\Psi_{\bs R}\rangle=0$. Therefore, only $N-1$ states are linearly independent. However, one can construct {\it two} additional eigenstates in real space which are linearly independent from the states in Eq.~\eqref{eq:PsiR}. They have support on non-contractible loops which encircle the torus in one or the other direction, as shown in Fig.~\ref{fig:tb_orbitals}(c). Therefore, the total number of linearly independent eigenstates with energy equal to the flat band is $N+1$. It follows that there must be an isolated point where the flat band touches the dispersing band.\cite{Wu:2007,Bergman:2008} The corresponding  ${\bs k}={\bs 0}$ Bloch states are given by
\begin{eqnarray}
|\Psi_{{\bs 0},x^2-y^2}^{\nu}\rangle&=&\frac{1}{\sqrt{\mathcal N}}\left(\sum_{i\in A}|d_{{\bs r},x^2-y^2}\rangle-\nu\sum_{i \in B}|d_{i,x^2-y^2}\rangle\right),\nonumber\\
|\Psi_{{\bs 0},z^2}^{\nu}\rangle&=&\frac{1}{\sqrt{\mathcal N}}\left(\sum_{i \in A}|d_{i,z^2}\rangle-\nu\sum_{i \in B}|d_{i,z^2}\rangle\right).
\end{eqnarray}
with $\mathcal{N}$ a normalization factor. This result from real-space topology is consistent with the band structure obtained by direct diagonalization.

\subsection{Generalized tight-binding model}
The idealized model Eq.~\eqref{eq:Ht} of the previous section only includes the hopping $t$ between the directed $d_{u^2}$-orbitals. To get a satisfying fit to the LDA band structure, the tight-binding model has to be generalized by including additional hopping terms. Although these terms are small they nevertheless modify the band structure notably. In particular, additional hopping amplitudes introduce a weak dispersion in the otherwise flat bands. It turns out that the two-fold degeneracy at ${\bs k}={\bs 0}$ is protected by the trigonal symmetry ($D_{3d}$) of the bilayer system and therefore is robust against the inclusion of additional tight-binding parameters. In fact, the two pairs of degenerated eigenstates at ${\bs k}={\bs 0}$ span the vector spaces of the two-dimensional irreducible representations of the $D_{3d}$ point group. This fact is interesting because if the Fermi surface coincides with the symmetry-protected quadratic band touching point, unusual weak coupling instabilities are possible.\cite{Sun:2009,Wen:2010,Ruegg:2011c,Uebelacker:2011} The symmetry-broken state generically is either a (lattice) nematic phase in which the rotational symmetry is broken or a topological phase with a gapped band structure characterized by a non-trivial topological invariant.

In the following we refine the tight-binding model by including additional parameters consistent with the $e_g$ character of the orbitals and assuming that the trigonal symmetry of the lattice is preserved.
\subsubsection{Nearest-neighbor hopping}
The Slater-Koster parameters for hopping along the $z$-direction yield the matrix
\begin{equation}
\hat{t}_z=-\begin{pmatrix}
t&0\\
0&t_{\delta}
\end{pmatrix}
\label{eq:tz}
\end{equation}
in the basis $(d_{z^2},d_{x^2-y^2})$. As before, $t$ includes predominantly the hopping via the intermediate oxygen while $t_{\delta}$ arises from the direct overlap and is small, see below. Assuming that the nearest-neighbor hopping in $x$ and $y$ directions are equivalent to the hopping along the $z$ direction, we obtain the corresponding matrices by a rotation of the $e_g$-orbitals around [111] by $\pm 2\pi/3$. The matrix for the rotation by $2\pi/3$ is
\begin{equation}
\hat{R}=\begin{pmatrix}
-1/2&\sqrt{3}/2\\
-\sqrt{3}/2&-1/2
\end{pmatrix}.
\label{eq:R}
\end{equation}
As a result, we then find
\begin{equation}
\hat{t}_x=\hat{R}^T\hat{t}_z\hat{R}, \quad \hat{t}_y=\hat{R}^T\hat{t}_x\hat{R}.
\end{equation}
\subsubsection{Second-neighbor hopping}
The Slater-Koster parameters for second-neighbor hopping via two intermediate oxygen atoms define the matrix
\begin{equation}
\hat{t}_{xy}=-\begin{pmatrix}
t'/2&\sqrt{3}\Delta/2\\
-\sqrt{3}\Delta/2&-3t'/2
\end{pmatrix}.
\label{eq:txy}
\end{equation}
They take into account the lowest-order processes for the second-neighbor hopping. The off-diagonal entries proportional to $\Delta$ are allowed in the bilayer system discussed here (as opposed to a perfect cubic system) because the two possible paths connecting second-neighbor transition-metal ions are not equivalent: they either involve ``inner" or ``outer" oxygens, see also Fig.~\ref{fig:NiOmodel}. Note that $t_{xy}$ is not symmetric if $\Delta\neq 0$ which means that there is an associated direction for the hopping. We use the convention that $t_{xy}$ denotes the hopping of an electron along a second neighbor bond which is reached by first following the $y$-axis and then the $x$-axis of the cube.
By rotating the orbitals, we also obtain the second-neighbor hopping along the other directions:
\begin{equation}
\hat{t}_{yz}=\hat{R}^T\hat{t}_{xy}\hat{R},\quad \hat{t}_{zx}=\hat{R}^T\hat{t}_{yz}\hat{R}.
\end{equation}
Including the above introduced hopping matrices, the generalized tight-binding model now takes the form
\begin{eqnarray}
H_0&=&\sum_{{\bs r}\in A}\sum_{\sigma}\sum_{u=xyz}\left(\vec{d}^{\dag}_{\sigma,{\bs r}}\hat{t}_u\vec{d}_{\sigma,{\bs r}+{\bs e}_u}+{\rm h.c.}\right)\nonumber\\
&&+\sum_{{\bs r}\in A}\sum_{\sigma}\sum_{u=xyz}\left(\vec{d}^{\dag}_{\sigma,{\bs r}}\hat{t}_{u,u+1}\vec{d}_{\sigma,{\bs r}+{\bs e}_u-{\bs e}_{u+1}}+{\rm h.c.}\right)\label{eq:H0}\\
&&+\sum_{{\bs r}\in B}\sum_{\sigma}\sum_{u=xyz}\left(\vec{d}^{\dag}_{\sigma,{\bs r}}\hat{t}_{u,u+1}\vec{d}_{\sigma,{\bs r}-{\bs e}_u+{\bs e}_{u+1}}+{\rm h.c.}\right).\nonumber
\end{eqnarray}
Here, $\vec{d}_{\sigma}=(d_{z^2,\sigma},d_{x^2-y^2,\sigma})^T$ is a vector in orbital space and the notation $u+1$ refers to $y$ if $u=x$ with a cyclic extension to the other elements. 

\subsection{Tight-binding fit of LDA band structure}
\label{sec:fit}
\begin{figure}
\includegraphics[width=0.9\linewidth]{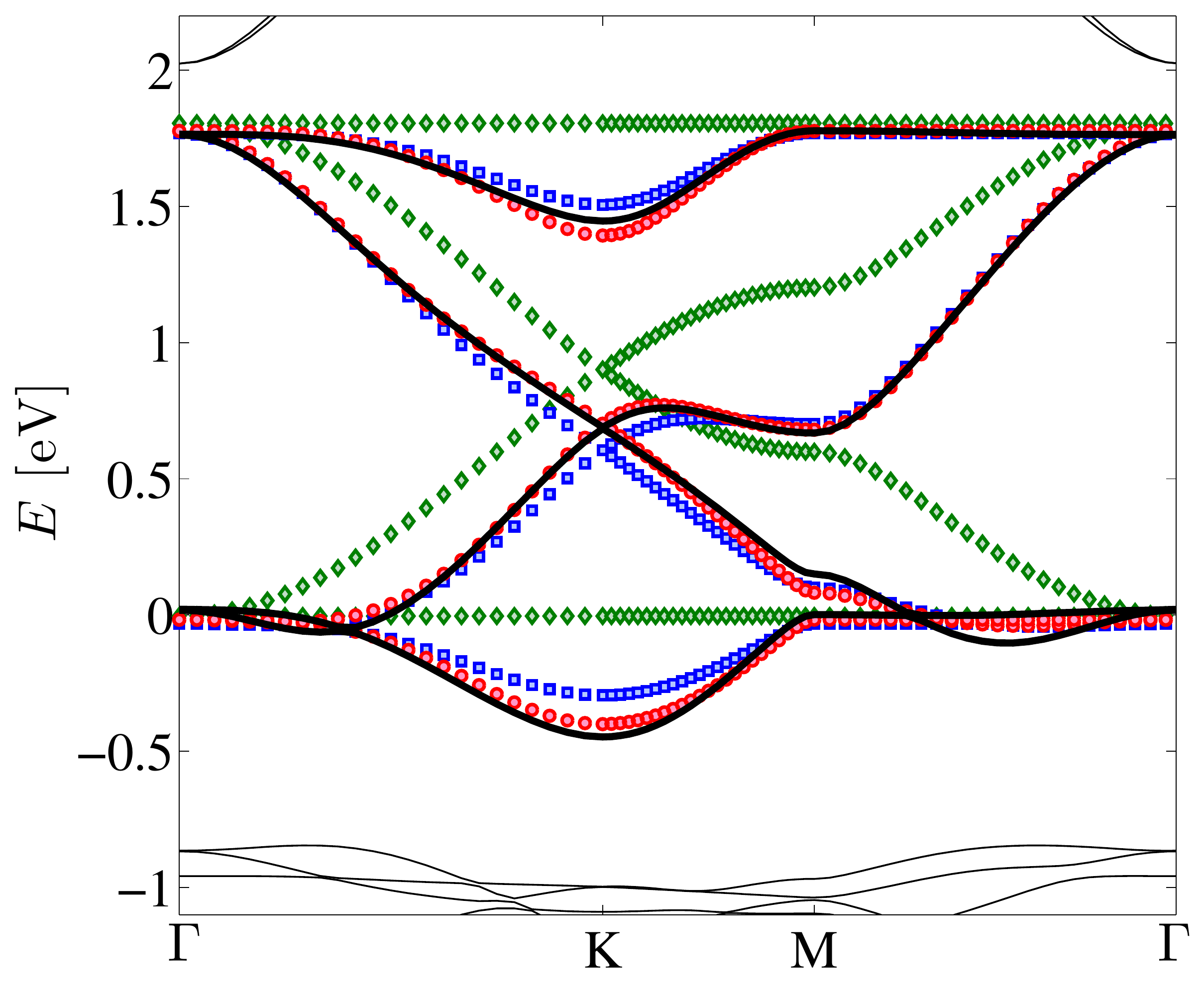}
\caption{(Color online). The solid line shows the LDA bands near the Fermi energy along a high-symmetry path $\Gamma$-K-M-$\Gamma$ for an unrelaxed (LaNiO$_3$)$_2$/(LaAlO$_3$)$_{10}$ superlattice. Also shown are the tight-binding bands obtained from a least-square fit as detailed in the main text and Tab.~\ref{tab:parameters}. Shown are the fits A (diamonds), B (squared) and C (circles) while fit D is essentially indistinguishable from C and not shown. The bands of A have been shifted by +0.2 eV for better visibility.}
\label{fig:band_fit}
\end{figure}

We used the refined tight-binding model Eq.~\eqref{eq:H0} for a least-square fit of the LDA band structure of the $e_g$-bands along the path $\Gamma$-K-M-$\Gamma$. The result is shown in Fig.~\ref{fig:band_fit} and the fitting parameters are listed in Tab.~\ref{tab:parameters}. We have performed fits with an increasing number of adjustable hopping amplitudes $t$, $t'$, $\Delta$ and $t_{\delta}$. We find that the nearest-neighbor hopping is about $t=0.6$ eV. The next biggest parameter is the second neighbor hopping for which we find $t'\approx0.1 t$. This values are in agreement with previous findings.\cite{Yang:2011,LeeS:2011b}  Even smaller is the off-diagonal term $\Delta$ in Eq.~\eqref{eq:txy} which is specific to the bilayer system and we find $\Delta\approx -0.04t$. However, the inclusion of this parameter markedly improves the fit, especially near the $K$-point, as can be seen in Fig.~\ref{fig:band_fit}. Finally, we find that the inclusion of the direct overlap $t_{\delta}$ is vanishingly small and does not lead to an essentially better fit.
\begin{table}
\begin{ruledtabular}
\begin{tabular}{l | l l l l l}
fit & $t$ [eV]& $t'$ [eV]& $\Delta$ [eV]& $t_{\delta}$ [eV] & $E_F$ [eV]\\
\hline
A &  0.603 & 0 & 0 & 0 & -0.701\\
B &  0.600 & 0.058 & 0 & 0& -0.693\\
C & 0.598 & 0.062 & -0.023 & 0 &  -0.693\\
D & 0.598 & 0.062 & -0.023 & -0.007 &  -0.693
\end{tabular}
\end{ruledtabular}
\caption{Parameters obtained in different tight-binding fits to the $e_g$ LDA band structure of the 12 layer superlattice shown in Fig.~\ref{fig:band_fit}.  with an increasing number of adjustable parameters.}
\label{tab:parameters}
\end{table}

We have also performed fits whit a tight-binding model with completely general hopping matrices for $t_z$ [Eq.~\eqref{eq:tz}] and $t_{xy}$ [Eq.~\eqref{eq:txy}]. However, we found that the least-square fit is only marginally better indicating that the most dominant processes are included in the form of the matrices in Eqs.~\eqref{eq:tz} and \eqref{eq:txy}.
Despite the good overall agreement, certain details of the LDA band structure (e.g. the low-lying part near M and along M-$\Gamma$) are not captured within the present model. This indicates that in order to describe all the details of the LDA band-structure in a $e_g$ tight-binding model, one has to consider further-range hopping parameters. An alternative way is to explicitly include the oxygen-$p$ states. This is done in Sec.~\ref{sec:CT} where we discuss a Ni-O lattice model. 

\subsection{Interaction effects in $e_g$-model}
We now turn to an analysis of interaction effects within the effective description of the tight-binding model Eq.~\eqref{eq:H0}. Owing to the localized character of the $3d$-orbitals, we include local interactions of the standard form\cite{Mizokawa:1996,Imada:1998}
\begin{eqnarray}
H_{\rm int}&&=\sum_{\bs r}\Big[U\sum_ {\alpha}n_{{\bs r}\alpha\uparrow}n_{{\bs r}\alpha\downarrow}+(U'-J)\sum_{\alpha>\beta,\sigma}n_{{\bs r}\alpha\sigma}n_{{\bs r}\beta\sigma}\nonumber\\
&&+U'\sum_{\alpha\neq \beta}n_{{\bs r}\alpha\uparrow}n_{{\bs r}\beta\downarrow}+J\sum_{\alpha\neq \beta}d_{{\bs r}\alpha\uparrow}^{\dag}d_{{\bs r}\beta\uparrow}d_{{\bs r}\beta\downarrow}^{\dag}d_{{\bs r}\alpha\downarrow}\nonumber\\
&&+I\sum_{\alpha\neq \beta}d_{{\bs r}\alpha\uparrow}^{\dag}d_{{\bs r}\beta\uparrow}d_{{\bs r}\alpha\downarrow}^{\dag}d_{{\bs r}\beta\downarrow}\Big].
\label{eq:Hint}
\end{eqnarray}
We assume the following relations between the Slater-Kanamori interaction parameters: $U'=U-2J$ and $I=J$. They are valid in free space and considered as approximately true in the solid state environment. The total multi-orbital Hubbard Hamiltonian for the $e_g$ electrons is given by
\begin{equation}
H=H_0+H_{\rm int}.
\label{eq:Htot}
\end{equation}
The interacting Hamiltonian Eq.~\eqref{eq:Htot} with only nearest-neighbor hopping has been studied previously within the Hartree-Fock approximation and the phase diagram has been worked out for various combinations of interaction parameters.\cite{Yang:2011,Ruegg:2011c} A particularly interesting result for intermediate to strong interactions is the observation of a spontaneously generated quantum anomalous Hall (QAH) phase which is accompanied by ordering in complex orbitals within a ferromagnetic (FM) phase. The resulting mean-field band structure is topologically non-trivial with a finite Chern number $\nu=\pm1$ displaying a spontaneous quantum Hall effect.

\begin{figure}
\includegraphics[width=0.95\linewidth]{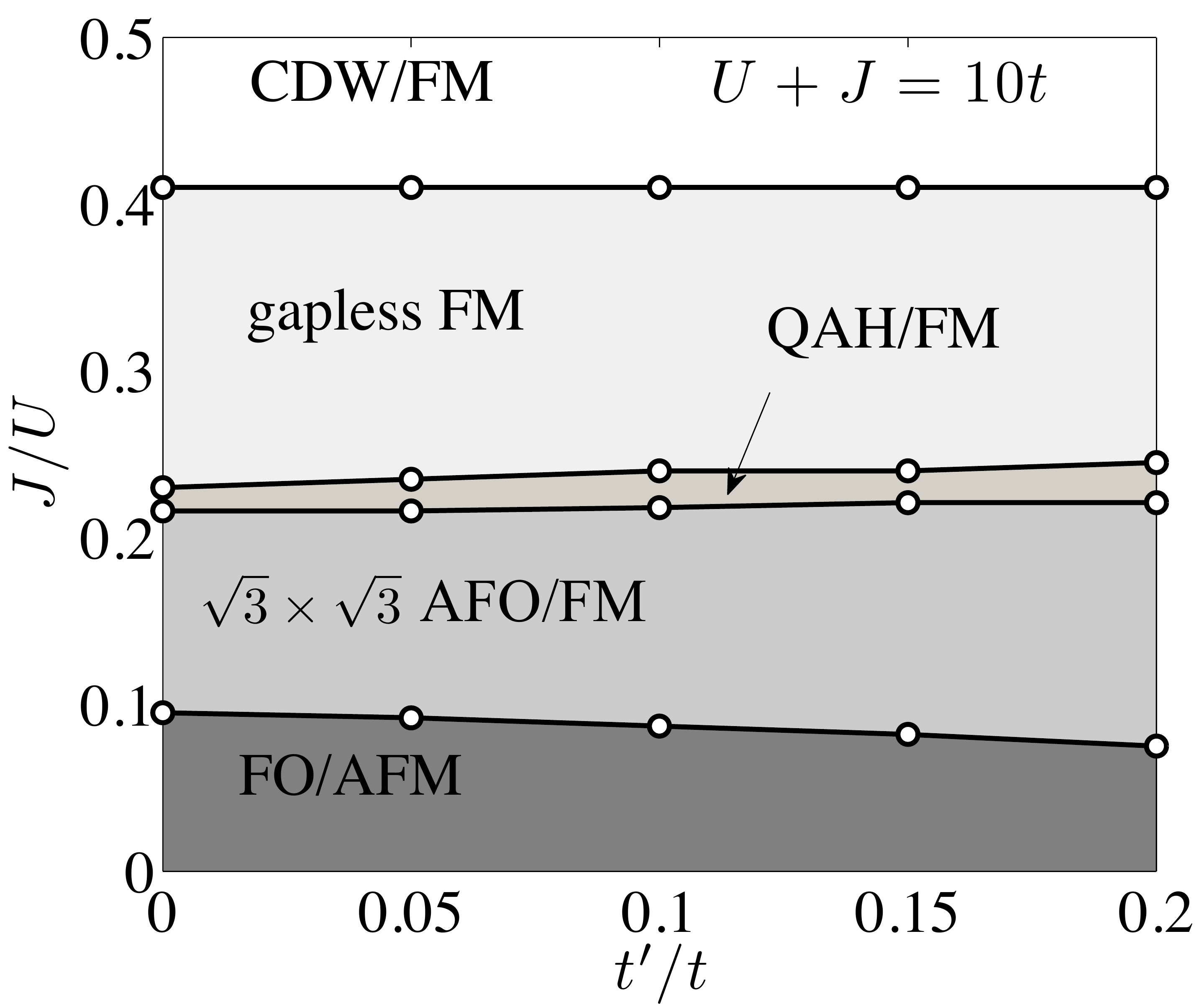}
\caption{Hartree-Fock phase diagram obtained from the interacting $e_g$-model for fixed interaction $U+J=10t$ as function of the ratio $J/U$ and $t'/t$. The various phases are explained in the main text.}
\label{fig:phasediagram}
\end{figure}

Here, we generalize the previous Hartree-Fock studies in three ways: We study the effect of a finite second-neighbor hopping $t'$, we allow for ordering with a tripled unit cell and we also consider larger values of the Hund coupling $J$. The resulting phase diagram for fixed interaction $U+J=10t$ as function of the ratio $J/U$ and $t'/t$ is shown in Fig.~\ref{fig:phasediagram}. Our new findings are summarized as follows: (i) For $0.1\lesssim J/U\lesssim 0.2$ we find that a ferromagnetic phase (FM) with staggered orbital order (AFO) which triples the unit cell but preserves the three-fold rotation symmetry has lower energy than the previously found phase which breaks the rotation symmetry but preserves the original unit cell.\cite{Yang:2011,Ruegg:2011c} (ii) For large Hund coupling $J/U\gtrsim 0.4$ a charge-density wave (CDW) is stabilized. (iii) For the considered interactions, there is very little dependence on $t'$. As a side note, we remark that the phase-boundaries at weak interactions, as discussed in Ref.~\onlinecite{Ruegg:2011c}, are more strongly affected by the inclusion of the second-neighbor hopping $t'$. In particular, upon increasing $t'/t$, we observe a rapid shrinking ($t'/t=0.01$ is enough) of the region where the spontaneous topological insulator phase is energetically favored.

In the following, we discuss the various phases shown in Fig.~\ref{fig:phasediagram} in more detail. For the considered range of parameters, we find ferromagnetic order of the spin degrees within a Ni-layer. However, depending on the relative strength of the Hund exchange, the coupling between the two layers is antiferromagnetic (AFM) if $J/U\lesssim 0.1$ or ferromagnetic (FM) if $J/U\gtrsim 0.1$. 

In addition to the magnetic order, we find various types of orbital order. The orbital order is conveniently discussed with the help of the orbital pseudospin-1/2 operator
\begin{equation}
\vec{T}_i=\frac{1}{2}\sum_{\sigma}\sum_{\alpha,\beta}d_{i\alpha\sigma}^{\dag}\vec{\tau}_{\alpha\beta}d_{i\beta\sigma}.
\end{equation}
The spontaneous development of long-range orbital order of the orbital pseudospin is indicated by a finite expectation value
\begin{equation}
\vec{\mu}_i=\langle \vec{T}_i\rangle.
\end{equation}
Depending on the details of the interaction parameters, we find different types of orbital order.
\subsubsection{FO/AFM}
For antiferromagnetically coupled layers, we found ferro-orbital (FO) order with 
\begin{equation}
\vec{\mu}=
\begin{pmatrix}
0\\0\\\mu_z
\end{pmatrix}.
\end{equation}
Other equivalent orbital orders are obtained by rotating the real orbitals around [111] by $\pm 2\pi/3$:
\begin{equation}
\begin{pmatrix}
\mu'_z\\\mu'_x
\end{pmatrix}
=
R
\begin{pmatrix}
\mu_z\\\mu_x
\end{pmatrix}.
\end{equation}
Here, the rotation matrix is given in Eq.~\eqref{eq:R}. Hence, the FO/AFM order breaks the rotation symmetry while preserving the translational symmetry.
\subsubsection{$\sqrt{3}\times\sqrt{3}$ AFO/FM}
Increasing $J/U$ we find a first-order transition to the ferromagnetic phases. In this regime, spin-up and spin-down bands are separated by an energy $(U+J)/2$ and we find fully spin-polarized phases. Because of the large energy splitting between $\uparrow$ and $\downarrow$ spins, the orbital order in this regime can be studied in a spinless model. Assuming polarization along the $\uparrow$-direction and neglecting the $\downarrow$-bands, the effective Hamiltonian in the fully polarized ferromagnet is
\begin{equation}
H_{\rm FM}=H_{0,\uparrow}+V\sum_in_{ia\uparrow}n_{ib\uparrow},
\label{eq:HFM}
\end{equation}
with $V=U'-J=U-3J$. This model is formally equivalent to a ``single-orbital" Hubbard model with ``spin-dependent" hopping and Hubbard interaction $V$. The model Eq.~\eqref{eq:HFM} should be studied at half filling where the Fermi energy crosses the Dirac points for $V=0$.

\begin{figure}
\includegraphics[width=0.8\linewidth]{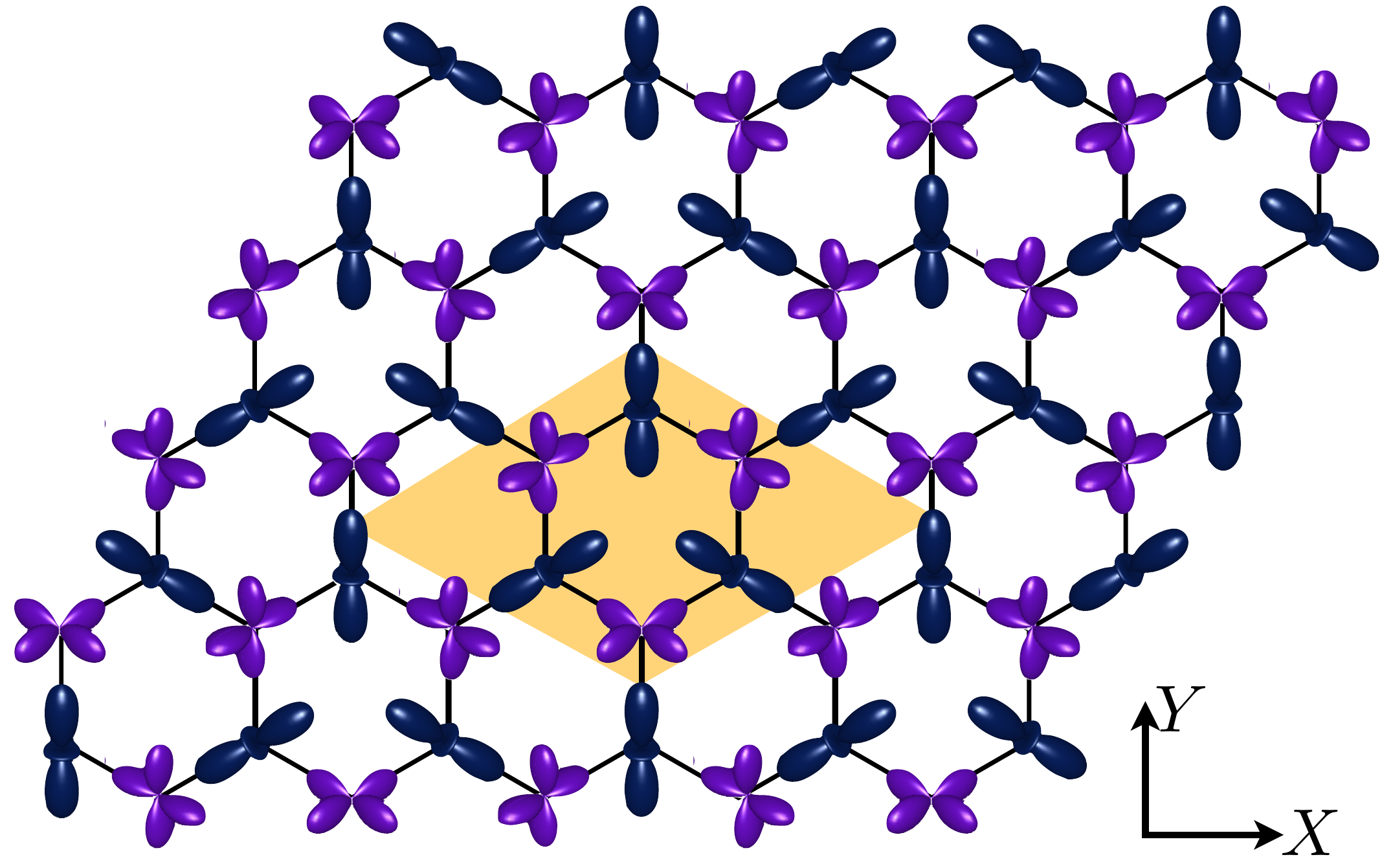}
\caption{(Color online.) Sketch of the staggered orbital order phase with a tripled unit cell ($\sqrt{3}\times\sqrt{3}$ AFO/FM). The shown real orbitals indicate the majority occupation and the enlarged unit cell is shown as shaded area.}
\label{fig:tripled}
\end{figure}

The $\sqrt{3}\times\sqrt{3}$ AFO/FM phase corresponds to the intermediate to large $V$ limit of Eq.~\eqref{eq:HFM}. It is characterized by a special type of staggered orbital order with a $\sqrt{3}\times\sqrt{3}$ reconstruction of the unit cell, see Fig.~\ref{fig:tripled}. In this phase, the orbital order parameters at different sites are related by a unitary transformation. For example, the order parameters at site 3 of Fig.~\ref{fig:tb_orbitals}(b) are related to site 1 in the following way:
\begin{equation}
\begin{pmatrix}
\mu_3^z\\\mu_3^x
\end{pmatrix}
=
R
\begin{pmatrix}
\mu_1^z\\\mu_1^x
\end{pmatrix}
\end{equation}
This corresponds to a rotation of the orbitals around the [111] axis by an angle of $2\pi/3$. A real space sketch of the orbital order in this phase is shown in Fig.~\ref{fig:tripled}. On the mean-field level, there is a {\it six}-fold degeneracy of this phase.

The findings of the mean-field theory can be compared with the strong coupling expansion of Eq.~\eqref{eq:HFM}. At half filling, Eq.~\eqref{eq:HFM} reduces to an orbital pseudospin model of the form
\begin{equation}
H_{K}=K\sum_{i\in A}\left(\tau_i^x\tau_{i+e_x}^x+\tau_i^y\tau_{i+e_y}^y+\tau_i^z\tau_{i+e_x}^z\right)
\label{eq:HK}
\end{equation}
where $K=2t^2/V$. In terms of the orbital pseudospin-1/2 operator $\vec{T}_i$,
the operators entering Eq.~\eqref{eq:HK} are given by
\begin{eqnarray}
\tau_i^x&=&-1/2T_i^z-\sqrt{3}/2T_i^x,\\
\tau_i^y&=&-1/2T_i^z+\sqrt{3}/2T_i^x,\\
\tau_i^z&=&T_i^z.
\end{eqnarray}
Because $K>0$, staggered orbital order is energetically favored, consistent with the mean-field result. Nevertheless, the pseudospin interaction described by Eq.~\eqref{eq:HK} is frustrated on the honeycomb lattice and to our knowledge the nature of the ground-state order hasn't been identified completely. Indeed, the study of the classical version of Eq.~\eqref{eq:HFM} in Ref.~\onlinecite{Nasu:2008} shows that the ground-state is macroscopically degenerate. Furthermore, it was shown that both thermal and quantum fluctuations lift this degeneracy. The $\sqrt{3}\times\sqrt{3}$ reconstruction identified in our mean-field analysis is insofar a reasonable scenario because it maximizes the number of ``resonating hexagons" favored by ring exchange processes. Such a behavior would be similar to the one observed in other frustrated models such as the hardcore dimer model on the hexagonal lattice\cite{Moessner:2001} or the frustrated charge models on the kagome\cite{OBrien:2010,Ruegg:2011} or checker board lattices.\cite{Pollmann:2006}

\subsubsection{QAH/FM}
In a narrow region of $J/U$, which corresponds to an intermediate effective interaction $V$ in the spinless model Eq.~\eqref{eq:HFM}, we find a phase which is characterized by the uniform orbital order along the $y$-direction, i.e. 
\begin{equation}
\mu_y\neq 0.
\end{equation}
This means that the electrons predominantly occupy complex orbitals
\begin{equation}
|d\pm id\rangle=\frac{1}{\sqrt{2}}\left(|d_{z^2}\rangle\pm i|d_{x^2-y^2}\rangle\right).
\end{equation}
Such orbital order opens a gap throughout the Brillouin zone. It has been pointed out previously\cite{Ruegg:2011c,Yang:2011} that the resulting insulator is a spontaneous quantum anomalous Hall (or Chern) insulator with topologically protected chiral edge modes. The mean-field band structure is characterized by a finite Chern number\cite{Thouless:1982} $n=\pm1$.
\subsubsection{Gapless FM}
Increasing the ratio $J/U$ even further, we find a gapless ferromagnetic phase without long-range orbital order. This phase corresponds to the weak-coupling phase of the spinless model Eq.~\eqref{eq:HFM} and it reflects the well-known fact that the Dirac semi-metal is perturbatively stable against interactions. Therefore, orbital order is suppressed and $\vec{\mu}=0$. Instead, the ferromagnetic solution with gapless single-particle excitations is stable.

\subsubsection{CDW/FM}
By further increasing the relative strength of the Hund coupling, we identify a second order phase transition to a charge-density wave (CDW) with charge disproportion between the top and bottom layer (i.e. between the A and B sublattices of the buckled honeycomb lattice). The tendency towards a CDW phase can be understood from the effective model Eq.~\eqref{eq:HFM} which shows that the effective interaction $V$ becomes {\it attractive} if $J>U/3$. Our mean-field analysis reveals that an even larger critical ratio $J/U\approx 0.4$ is needed to stabilize the charge disproportion. 

A CDW with charge disproportion between the A and B sublattices has also been proposed for bulk nickelates as a mechanism to lift the orbital degeneracy which is different from the Jahn-Teller distortion.\cite{Mazin:2007} Our findings that a relatively large Hund coupling is required is in agreement with a recent work based on a two-band model for bulk LaNiO$_3$.\cite{LeeS:2011b}

\section{Nickel-Oxygen model}
\label{sec:CT}
In the previous section we discussed the band structure of the conduction electrons in terms of an effective multi-orbital Hubbard model for the $e_g$ states. In this model, the oxygens enter the description implicitly by mediating the hopping between the Ni ions. The satisfactory fit of the LDA band structure from the tight-binding model obtained in Sec.~\ref{sec:fit} seems to justify this view-point. On the basis of this model, we have obtained the Hartree-Fock phase diagram which, depending on the interaction parameters, shows both orbitally ordered and disordered phases. While the ferromagnetic order predicted by the Hartree-Fock for a large region of parameters is consistent with the LSDA$+U$ result, we did not observe orbital ordering in the DFT calculations. 

Below we reexamine the stability of the orbitally ordered phases in the context of a lattice model which explicitly includes both the oxygen and nickel ions. Within the Hartree-Fock approximation, we study the dependence of the electronic phases on the energy splitting between the Ni-$d$ and O-$p$ states and find that orbital order is more stable if the charge-transfer energy is increased. We discuss possible mechanisms leading to this result.

\subsection{Ni-O model for (111) bilayer}
\begin{figure}
\includegraphics[width=0.6\linewidth]{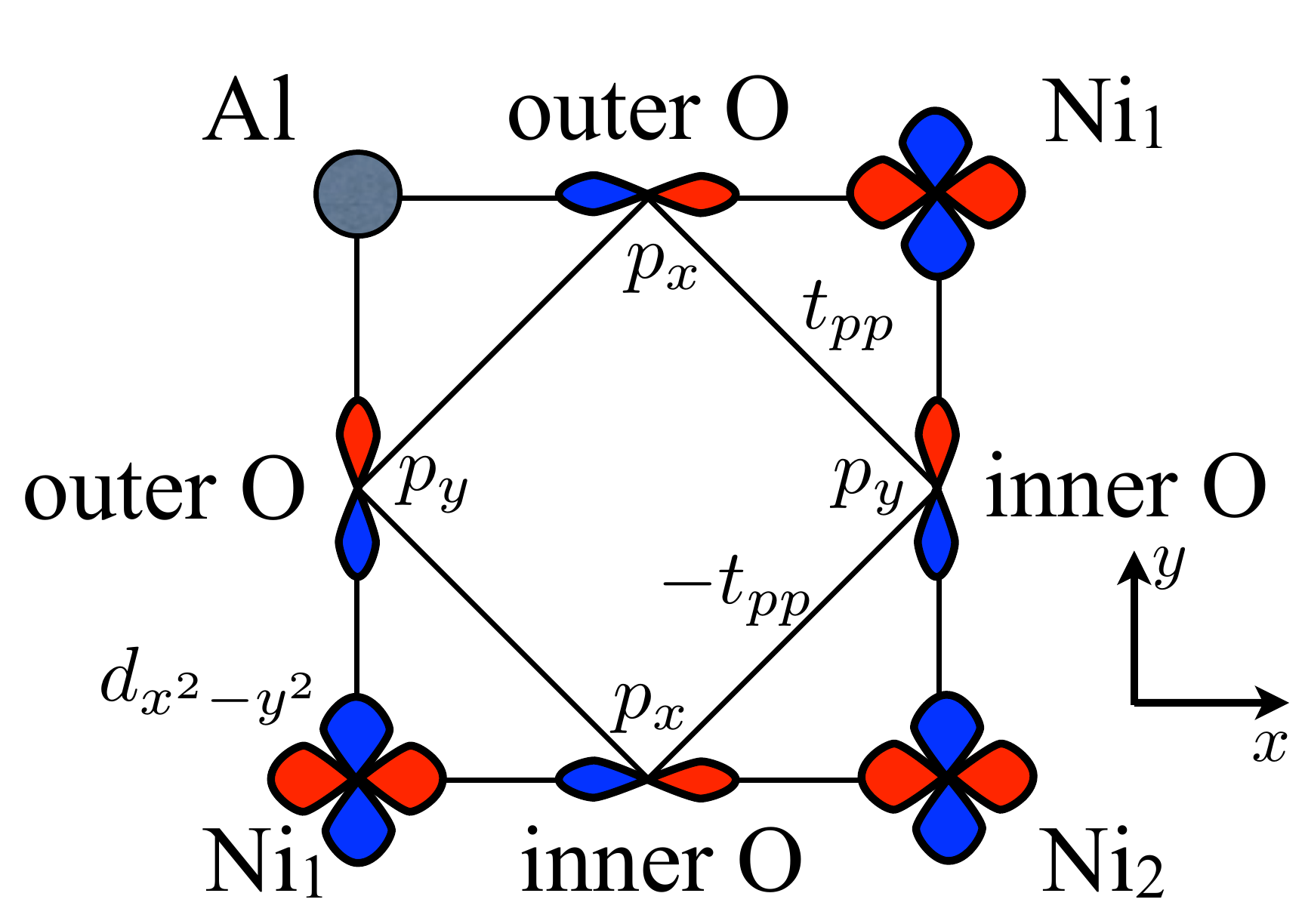}
\caption{(Color online.) Sketch of the Ni-O model for the (111) bilayer. There are two layers of Ni ions denoted by Ni$_1$ and Ni$_2$ for which we keep the $e_g$ orbitals. There are an inner and two outer layers of O ions. We keep the O-$p$ orbitals which form a $\sigma$-bond with their neighboring Ni-$e_g$ orbitals. The Al ions are treated as vacuum, so that there is no mixing with its neighboring oxygens.}
\label{fig:NiOmodel}
\end{figure}

In the following, we consider a model which treats both the Ni-$d$ and O-$p$ orbitals.\cite{Mizokawa:1996} Instead of keeping the full Ni-$d$ and O-$p$ manifolds, we consider a simplified version with two $e_g$-orbitals at the Ni sites and one $p$-orbital at the O sites, see Fig.~\ref{fig:NiOmodel}. The Al ions are treated as vacuum. The $p$ orbital is chosen such that it points along the directions of the cartesian coordinate system, i.e. the $p$-orbital which makes a $\sigma$-bond with its neighboring $d_{x^2-y^2}$/$d_{3z^2-r^2}$ orbitals. The resulting model has the form
\begin{eqnarray}
H&=&\sum_{i\sigma}\epsilon_{pi} p_{i\sigma}^{\dag}p_{i\sigma}+\sum_{i\alpha\sigma}\epsilon_d d_{i\alpha\sigma}^{\dag}d_{i\alpha\sigma}+H_{\rm hyb}+H_{p}\nonumber\\
&&+H_{\rm int}+H_{DC}.
\label{eq:Hpd}
\end{eqnarray}
The hybridization between the $p$ and $d$ electrons is given by
\begin{equation}
H_{\rm hyb}=\sum_{\langle i,j\rangle}\left(V_{ij}^{\alpha}p_{i\sigma}^{\dag}d_{j\alpha\sigma}+{\rm h.c.}\right).
\end{equation}
The hybridization $V_{ij}^{\alpha}$ between the $e_g$ orbitals and the $p$ orbitals is parametrized by the Slater-Koster parameter $(pd\sigma)$.
$H_{p}$ describes the direct overlap between O-$p$ orbitals for which we introduce the hopping parameter $t_{pp}$. Finally, only the $d$-electrons are assumed to be correlated and $H_{\rm int}$ is given by Eq.~\eqref{eq:Hint}. $H_{DC}$ accounts for the ``double-counting" in the interacting model and will be discussed later.

The unit cell of the bilayer system has two nickel and nine oxygen atoms arranged in layers along the [111] direction as O$_3$/Ni/O$_3$/Ni/O$_3$. By symmetry, the inner oxygen layer, which is sandwiched between the Ni layers, has a different onsite-energy $\epsilon_{p}^{(i)}$ than the outer oxygens $\epsilon_{p}^{(o)}$.

\subsection{LDA fit based on Ni-O model}
\begin{figure}
\includegraphics[width=1\linewidth]{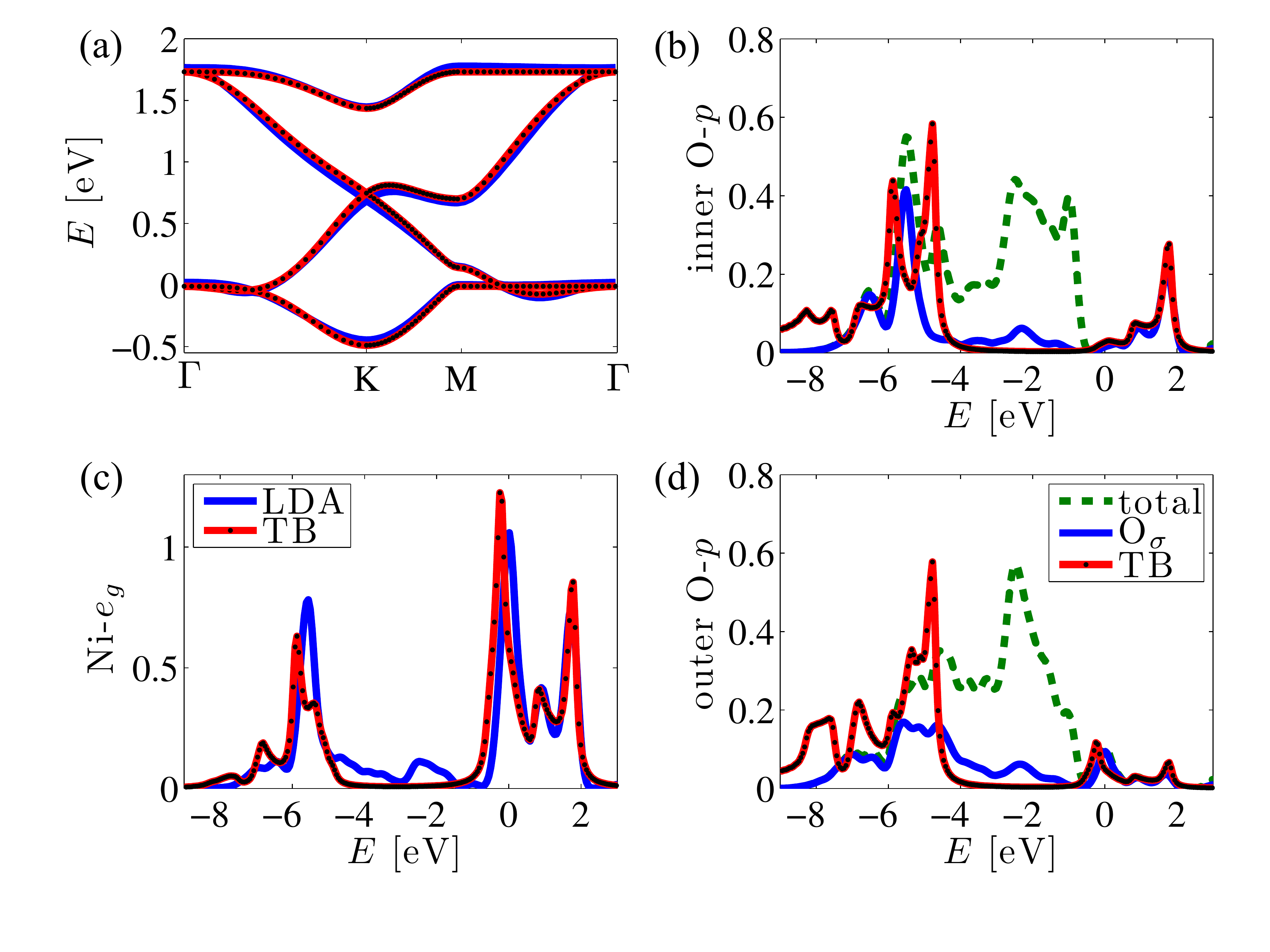}
\caption{(Color online.) Comparison between the LDA and the Ni-O tight-binding (TB) model. In (a) we show the least-square fit of the $e_g$ band structure along a high symmetry line. Panels (b)-(d) show the projected density of states obtained from the TB and the LDA calculations. The Ni-$e_g$ states are shown in (c) and the inner and outer O-$p$ states in (b) and (d). Along with the total O-$p$ DOS we also show the contribution from the O$_{\sigma}$ orbital alone.}
\label{fig:pdfit}
\end{figure}

We find that there is an ambiguity in fitting the parameters of the Ni-O model to the LDA results. In particular, fitting different quantities accessible in the Ni-O model, such as the $e_g$ band structure, the orbitally projected DOS or any combination of these observables result in different optimized tight-binding parameters. Nevertheless, different fits which are in agreement with the overall structure of {\em both} the O and Ni states yield similar parameters. Our qualitative conclusions are therefore not affected by the details of the fitting procedure. Here, we show the result obtained by fixing the Slater-Koster parameters at $(pd\sigma)=1.8$ eV and the hopping between oxygens at $t_{pp}=0.7$~eV. We choose to fit the $e_g$ band structure by optimizing the onsite energies for the Ni states, $\varepsilon_d$, and for the inner and outer O states, $\varepsilon_p^{(o)}$ and $\varepsilon_p^{(i)}$. The fixed hopping parameters are similar to the ones used by other groups,\cite{Mizokawa:1996,Han:2010} and we find that this choice gives satisfactory agreement both for the $e_g$ bands as well as for the projected DOS. The optimized values are $\epsilon_d=-1.47$~eV, $\epsilon_{p}^{(i)}=-4.74$~eV and $\epsilon_{p}^{(o)}=-5.47$~eV. The result of this fit is summarized in Fig.~\ref{fig:pdfit} where we show the $e_g$ bands and the projected DOS. 

\subsection{Hartree-Fock approximation and double counting}
The Hartree-Fock decoupling of $H_{\rm int}$ [Eq.~\eqref{eq:Hint}] yields a term
\begin{equation}
H_c=\bar{U}\sum_{i}\langle n_d\rangle n_i-\frac{\bar{U}}{2}\sum_{i}\langle n_d\rangle^2
\label{eq:Hc}
\end{equation}
where $\langle n_d\rangle$ is the averaged occupation of the Ni states and $\bar{U}=(3U-5J)/4$. In general, $H_c$ leads to a shift of the $d$-orbital energy as compared to the O-$p$ levels which would modify the average occupations of the $d$-orbitals when interactions are included. In the LSDA$+U$ method,\cite{Anisimov:1997} one therefore adds a double counting term which essentially compensate $H_c$. It is argued that the DFT calculation already includes a local term similar to Eq.~\eqref{eq:Hc} and that the overall charge distribution (although not the orbital occupation) is obtained reliably from the DFT. The interacting Hamiltonian with the double-counting correction should therefore reproduce the DFT result in the absence of interaction-driven orbital order. This so-called double-counting problem appears also in the DFT+DMFT approach.\cite{Kotliar:2006} 
In the spirit of the LSDA$+U$ we choose the double-counting term $H_{DC}$ such that it compensates for the above discussed Hartree shift, $H_{DC}=-H_c$. The remaining interactions are treated in the conventional Hartree-Fock approximation and we solved the self-consistency equations via iteration.

\begin{figure}
\includegraphics[width=1\linewidth]{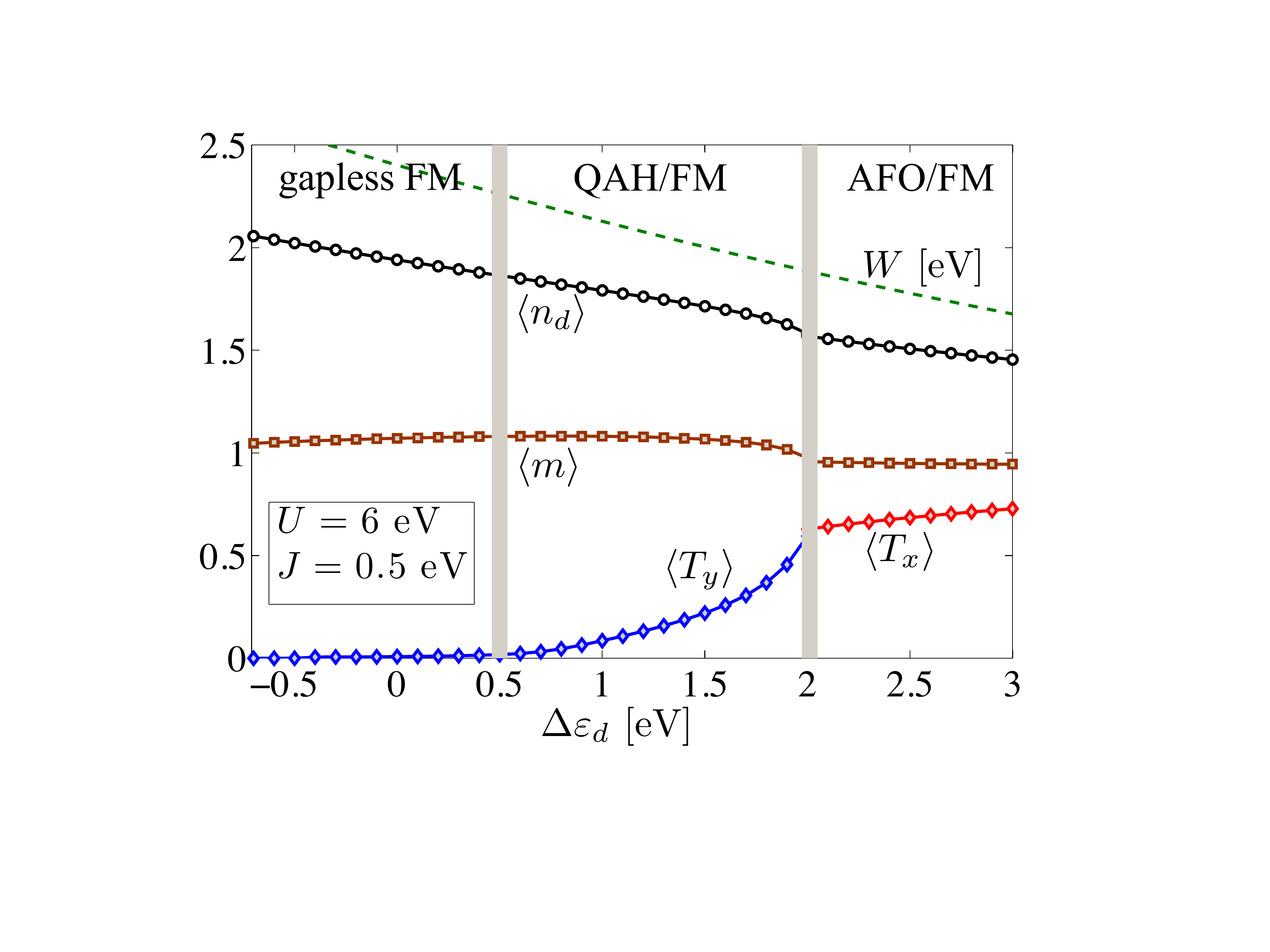}
\caption{Dependence of the symmetry-broken phases on the shift in the $d$-electron level $\Delta\varepsilon_d$ as obtained from the interacting Ni-O model in the Hartree-Fock approximation. Shown are the average occupation of the Ni-$e_g$ orbitals $\langle n_d\rangle$, the uniform magnetization $\langle m\rangle$ of the ferromagnetic (FM) order as well as the expectation values of the orbital pseudospin $\langle \vec{T}\rangle$. Increasing $\Delta\varepsilon_d$ triggers a transition from the gapless FM phase to a gapped topological phase (QAH) and eventually to a phase with staggered orbital order (AFO). Also shown is the bandwidth $W$ (in eV) of the $e_g$ band obtained in the corresponding non-interacting Ni-O model.}
\label{fig:Deltascan}
\end{figure}
\subsection{Charge transfer}
In order to see how the symmetry-broken phases are affected by the charge-transfer physics,\cite{Zaanen:1985,Mizokawa:2000,Khomskii:2001} we performed Hartree-Fock calculations for different charge-transfer energies using the double-counting term discussed in the previous section. We fixed the tight-binding parameters at the values obtained from the LDA except for $\varepsilon_d$ which we varied according to $\varepsilon_d\rightarrow\varepsilon_d+\Delta\varepsilon_d$. This effectively changes the energy splitting between the O and Ni states. Figure~\ref{fig:Deltascan} shows the dependence of the symmetry-broken phases on $\Delta\varepsilon_d$ for fixed interaction parameters $U=6$ eV and $J=0.5$ eV. Around $\Delta\varepsilon_d=0$, we find that a gapless ferromagnetic phase without orbital order is energetically favored. This is consistent with the LSDA$+U$ results reported in Sec.~\ref{sec:DFT}. Increasing $\Delta\varepsilon_d=0$ leads to orbital ordering of complex orbitals and $\langle T_y\rangle\neq0$ (QAH/FM). In this phase, the single-particle spectrum is gapped but the band-structure is topologically non-trivial and characterized by a finite Chern number. By increasing $\Delta\varepsilon_d$ even further, we find that staggered orbital order (AFO) in real orbitals is favored and $\langle T_x\rangle\neq 0$ on each sublattice.

There are two aspects of the charge-transfer physics which can help understanding these results. First, increasing the charge-transfer energy reduces the effective bandwidth of the $e_g$ bands. We illustrate this point in Fig.~\ref{fig:Deltascan} by plotting the width $W$ of the $e_g$ bands in the corresponding non-interacting Ni-O tight-binding model. The ratio $U/W$ is increased by increasing $\Delta\varepsilon_d$ and correlation effects become more important. The observed orbital order can therefore be considered as a result of increasing $U/W$ in an effective model for the $e_g$-electrons.\cite{Ruegg:2011c,Yang:2011} A second aspect of the charge-transfer physics can be quantified by the average occupation of the Ni-$e_g$ states, $\langle n_d\rangle$, which is also modified by the charge-transfer energy. For $\Delta\varepsilon_d=0$, the Ni-O model predicts $\langle n_d\rangle\approx 2$ which is very different from the ionic limit $\langle n_d\rangle=1$ (corresponding to Ni$^{3+}$) and indicates that a large number of holes resides on the oxygen sites. In our LSDA$+U$ calculation, the difference from the ionic value seems even larger: examining the local density-matrix we find $\langle n_d\rangle_{\rm DFT}\approx 2.4$.  Previously, this effect was called ``self-doping" of the oxygen $p$-band\cite{Mizokawa:2000,Khomskii:2001} and is considered to capture aspects beyond modifying the ratio of $U/W$ in an effective model for the $e_g$ electrons. Recently, it has been argued that $\langle n_d\rangle$ is a fundamental quantity which crucially affects the interaction-driven metal-insulator transition in the nickelates.\cite{Wangb:2011} In Fig.~\ref{fig:Deltascan}, the reduction in $\langle n_d\rangle$ goes hand in hand with the reduction of $W$ and therefore, these two aspects can not be separated in a clear way. Nevertheless, our results are consistent with the statement that the (gapped) orbitally ordered phases are more stable if fewer charges are transferred between the Ni and O. The charge transfer therefore offers an explanation why the orbitally ordered phases are not observed in our LSDA$+U$ calculation.

\section{Conclusions}
In summary, we discussed the electronic structure of a LaNiO$_3$ (111) bilayer sandwiched between several layers of the insulator LaAlO$_3$. Using a combination of first principle methods and effective multi-orbital lattice models we have studied both the non-interacting band structure as well as possible instabilities driven by local correlations among the Ni-$d$ electrons. The non-interacting band structure is well reproduced by a generalized tight-binding model including also second-neighbor hopping amplitudes for $e_g$ states on the buckled honeycomb lattice. If local interactions among the Ni-$d$ electrons are included, both the Hartree-Fock and the LSDA$+U$ predict ferromagnetic ordering over a wide range of parameters. We have extended previous Hartree-Fock calculations\cite{Yang:2011,Ruegg:2011c} in various ways and, apart from the magnetic order, we have also identified possible orbital ordering which could give rise to interesting electronic phases including a spontaneous Chern insulator or a staggered orbitally ordered phase with a triplet unit cell. Furthermore, we identified a charge-density wave for large values of the Hund coupling.

By considering an effective lattice model which includes both the O-$p$ and Ni-$d$ states, we studied the effect of modifying the charge-transfer energy between the Ni and O states on the orbital ordering. Our results are consistent with the statement that orbital ordering is suppressed the more charge is transferred. This is in qualitative agreement with recent theoretical results on the orbital polarization in [001] heterostructures\cite{Han:2010,Han:2011} and the phase boundaries for the paramagnetic metal-insulator transition in bulk nickelates and cuprates.\cite{Wangb:2011} The charge-transfer energy was experimentally observed to change with $n$ in (LaAlO$_3$)$_3$/(LaNiO$_3$)$_n$ superlattices\cite{Liu:2011} suggesting that it might be possible to control the degree of covalency in artificial structures. \cite{Chakhalian:2011} Important issues such as the atomic relaxation in the bilayer geometry and its dependence on the electronic correlations as well as the treatment of the electronic correlations beyond the static mean-field approximations remain topics for future study.

\begin{acknowledgements}
We acknowledge discussions with A.~J.~Millis, J.~Chakhalian and C.~Lin. G.A.F and A.R. acknowledge financial support through ARO grant W911NF-09-1-0527 and NSF grant DMR-0955778. A.A.D. and C.M. acknowledge financial support by the National Science Foundation under grant DMR-0548182, the US Department of Energy (DOE) under grant DE-SC0001878, and Texas Advanced Computing Center.
\end{acknowledgements}

\bibliography{biblio}

%Merlin.mbs v4.21 2009-07-09.
\begin{thebibliography}{10}%
\makeatletter
\providecommand \@ifxundefined [1]{%
 \ifx #1\undefined \expandafter \@firstoftwo
 \else \expandafter \@secondoftwo
\fi
}%
\providecommand \@ifnum [1]{%
 \ifnum #1\expandafter \@firstoftwo
 \else \expandafter \@secondoftwo
\fi
}%
\providecommand \enquote [1]{``#1''}%
\providecommand \bibnamefont  [1]{#1}%
\providecommand \bibfnamefont [1]{#1}%
\providecommand \citenamefont [1]{#1}%
\providecommand\href[0]{\@sanitize\@href}%
\providecommand\@href[1]{\endgroup\@@startlink{#1}\endgroup\@@href}%
\providecommand\@@href[1]{#1\@@endlink}%
\providecommand \@sanitize [0]{\begingroup\catcode`\&12\catcode`\#12\relax}%
\@ifxundefined \pdfoutput {\@firstoftwo}{%
 \@ifnum{\z@=\pdfoutput}{\@firstoftwo}{\@secondoftwo}%
}{%
 \providecommand\@@startlink[1]{\leavevmode\special{html:<a href="#1">}}%
 \providecommand\@@endlink[0]{\special{html:</a>}}%
}{%
 \providecommand\@@startlink[1]{%
  \leavevmode
  \pdfstartlink
   attr{/Border[0 0 1 ]/H/I/C[0 1 1]}%
   user{/Subtype/Link/A<</Type/Action/S/URI/URI(#1)>>}%
  \relax
 }%
 \providecommand\@@endlink[0]{\pdfendlink}%
}%
\providecommand \url  [0]{\begingroup\@sanitize \@url }%
\providecommand \@url [1]{\endgroup\@href {#1}{\urlprefix}}%
\providecommand \urlprefix [0]{URL }%
\providecommand \Eprint[0]{\href }%
\@ifxundefined \urlstyle {%
  \providecommand \doi [1]{doi:\discretionary{}{}{}#1}%
}{%
  \providecommand \doi [0]{doi:\discretionary{}{}{}\begingroup
  \urlstyle{rm}\Url }%
}%
\providecommand \doibase [0]{http://dx.doi.org/}%
\providecommand \Doi[1]{\href{\doibase#1}}%
\providecommand \bibAnnote [3]{%
  \BibitemShut{#1}%
  \begin{quotation}\noindent
    \textsc{Key:}\ #2\\\textsc{Annotation:}\ #3%
  \end{quotation}%
}%
\providecommand \bibAnnoteFile [2]{%
  \IfFileExists{#2}{\bibAnnote {#1} {#2} {\input{#2}}}{}%
}%
\providecommand \typeout [0]{\immediate \write \m@ne }%
\providecommand \selectlanguage [0]{\@gobble}%
\providecommand \bibinfo [0]{\@secondoftwo}%
\providecommand \bibfield [0]{\@secondoftwo}%
\providecommand \translation [1]{[#1]}%
\providecommand \BibitemOpen[0]{}%
\providecommand \bibitemStop [0]{}%
\providecommand \bibitemNoStop [0]{.\EOS\space}%
\providecommand \EOS [0]{\spacefactor3000\relax}%
\providecommand \BibitemShut [1]{\csname bibitem#1\endcsname}%
%</preamble>
\bibitem{Zubko:2011}%
  \BibitemOpen
  \bibfield{author}{%
  \bibinfo {author} {\bibfnamefont{P.}~\bibnamefont{Zubko}}, \bibinfo {author}
  {\bibfnamefont{S.}~\bibnamefont{Gariglio}}, \bibinfo {author}
  {\bibfnamefont{M.}~\bibnamefont{Gabay}}, \bibinfo {author}
  {\bibfnamefont{P.}~\bibnamefont{Ghosez}},\ and\ \bibinfo {author}
  {\bibfnamefont{J.-M.}\ \bibnamefont{Triscone}},\ }%
  \bibfield{journal}{%
  \Doi{10.1146/annurev-conmatphys-062910-140445}{\bibinfo {journal} {Annu. Rev.
  Condens. Matter Phys.}}\ }%
  \textbf{\bibinfo {volume} {2}},\ \bibinfo {pages} {141} (\bibinfo {year}
  {2011})%
  \bibAnnoteFile{NoStop}{Zubko:2011}%
\bibitem{Chakhalian:2012}%
  \BibitemOpen
  \bibfield{author}{%
  \bibinfo {author} {\bibfnamefont{J.}~\bibnamefont{Chakhalian}}, \bibinfo
  {author} {\bibfnamefont{A.~J.}\ \bibnamefont{Millis}},\ and\ \bibinfo
  {author} {\bibfnamefont{J.}~\bibnamefont{Rondinelli}},\ }%
  \bibfield{journal}{%
  \Doi{10.1038/nmat3225}{\bibinfo {journal} {Nat Mater}}\ }%
  \textbf{\bibinfo {volume} {11}},\ \bibinfo {pages} {92} (\bibinfo {year}
  {2012})%
  \bibAnnoteFile{NoStop}{Chakhalian:2012}%
\bibitem{Mannhart:2010}%
  \BibitemOpen
  \bibfield{author}{%
  \bibinfo {author} {\bibfnamefont{J.}~\bibnamefont{Mannhart}}\ and\ \bibinfo
  {author} {\bibfnamefont{D.~G.}\ \bibnamefont{Schlom}},\ }%
  \bibfield{journal}{%
  \Doi{10.1126/science.1181862}{\bibinfo {journal} {Science}}\ }%
  \textbf{\bibinfo {volume} {327}},\ \bibinfo {pages} {1607} (\bibinfo {year}
  {2010})%
  \bibAnnoteFile{NoStop}{Mannhart:2010}%
\bibitem{Ohtomo:2002}%
  \BibitemOpen
  \bibfield{author}{%
  \bibinfo {author} {\bibfnamefont{A.}~\bibnamefont{Ohtomo}}, \bibinfo {author}
  {\bibfnamefont{D.~A.}\ \bibnamefont{Muller}}, \bibinfo {author}
  {\bibfnamefont{J.~L.}\ \bibnamefont{Grazul}},\ and\ \bibinfo {author}
  {\bibfnamefont{H.~Y.}\ \bibnamefont{Hwang}},\ }%
  \bibfield{journal}{%
  \Doi{10.1038/nature00977}{\bibinfo {journal} {Nature}}\ }%
  \textbf{\bibinfo {volume} {419}},\ \bibinfo {pages} {378} (\bibinfo {year}
  {2002})%
  \bibAnnoteFile{NoStop}{Ohtomo:2002}%
\bibitem{Okamoto:2004a}%
  \BibitemOpen
  \bibfield{author}{%
  \bibinfo {author} {\bibfnamefont{S.}~\bibnamefont{Okamoto}}\ and\ \bibinfo
  {author} {\bibfnamefont{A.~J.}\ \bibnamefont{Millis}},\ }%
  \bibfield{journal}{%
  \Doi{10.1038/nature02450}{\bibinfo {journal} {Nature}}\ }%
  \textbf{\bibinfo {volume} {428}},\ \bibinfo {pages} {630} (\bibinfo {year}
  {2004})%
  \bibAnnoteFile{NoStop}{Okamoto:2004a}%
\bibitem{Chakhalian:2007}%
  \BibitemOpen
  \bibfield{author}{%
  \bibinfo {author} {\bibfnamefont{J.}~\bibnamefont{Chakhalian}}, \bibinfo
  {author} {\bibfnamefont{J.~W.}\ \bibnamefont{Freeland}}, \bibinfo {author}
  {\bibfnamefont{H.-U.}\ \bibnamefont{Habermeier}}, \bibinfo {author}
  {\bibfnamefont{G.}~\bibnamefont{Cristiani}}, \bibinfo {author}
  {\bibfnamefont{G.}~\bibnamefont{Khaliullin}}, \bibinfo {author}
  {\bibfnamefont{M.}~\bibnamefont{van Veenendaal}},\ and\ \bibinfo {author}
  {\bibfnamefont{B.}~\bibnamefont{Keimer}},\ }%
  \bibfield{journal}{%
  \Doi{10.1126/science.1149338}{\bibinfo {journal} {Science}}\ }%
  \textbf{\bibinfo {volume} {318}},\ \bibinfo {pages} {1114} (\bibinfo {year}
  {2007})%
  \bibAnnoteFile{NoStop}{Chakhalian:2007}%
\bibitem{Xiao:2011}%
  \BibitemOpen
  \bibfield{author}{%
  \bibinfo {author} {\bibfnamefont{D.}~\bibnamefont{Xiao}}, \bibinfo {author}
  {\bibfnamefont{W.}~\bibnamefont{Zhu}}, \bibinfo {author}
  {\bibfnamefont{Y.}~\bibnamefont{Ran}}, \bibinfo {author}
  {\bibfnamefont{N.}~\bibnamefont{Nagaosa}},\ and\ \bibinfo {author}
  {\bibfnamefont{S.}~\bibnamefont{Okamoto}},\ }%
  \bibfield{journal}{%
  \Doi{10.1038/ncomms1602}{\bibinfo {journal} {Nat Commun}}\ }%
  \textbf{\bibinfo {volume} {2}},\ \bibinfo {pages} {596} (\bibinfo {year}
  {2011})%
  \bibAnnoteFile{NoStop}{Xiao:2011}%
\bibitem{Yang:2011}%
  \BibitemOpen
  \bibfield{author}{%
  \bibinfo {author} {\bibfnamefont{K.-Y.}\ \bibnamefont{Yang}}, \bibinfo
  {author} {\bibfnamefont{W.}~\bibnamefont{Zhu}}, \bibinfo {author}
  {\bibfnamefont{D.}~\bibnamefont{Xiao}}, \bibinfo {author}
  {\bibfnamefont{S.}~\bibnamefont{Okamoto}}, \bibinfo {author}
  {\bibfnamefont{Z.}~\bibnamefont{Wang}},\ and\ \bibinfo {author}
  {\bibfnamefont{Y.}~\bibnamefont{Ran}},\ }%
  \bibfield{journal}{%
  \Doi{10.1103/PhysRevB.84.201104}{\bibinfo {journal} {Phys. Rev. B}}\ }%
  \textbf{\bibinfo {volume} {84}},\ \bibinfo {pages} {201104} (\bibinfo {year}
  {2011})%
  \bibAnnoteFile{NoStop}{Yang:2011}%
\bibitem{WangFa:2011}%
  \BibitemOpen
  \bibfield{author}{%
  \bibinfo {author} {\bibfnamefont{F.}~\bibnamefont{Wang}}\ and\ \bibinfo
  {author} {\bibfnamefont{Y.}~\bibnamefont{Ran}},\ }%
  \bibfield{journal}{%
  \Doi{10.1103/PhysRevB.84.241103}{\bibinfo {journal} {Phys. Rev. B}}\ }%
  \textbf{\bibinfo {volume} {84}},\ \bibinfo {pages} {241103} (\bibinfo {year}
  {2011})%
  \bibAnnoteFile{NoStop}{WangFa:2011}%
\bibitem{Ruegg:2011c}%
  \BibitemOpen
  \bibfield{author}{%
  \bibinfo {author} {\bibfnamefont{A.}~\bibnamefont{R\"uegg}}\ and\ \bibinfo
  {author} {\bibfnamefont{G.~A.}\ \bibnamefont{Fiete}},\ }%
  \bibfield{journal}{%
  \Doi{10.1103/PhysRevB.84.201103}{\bibinfo {journal} {Phys. Rev. B}}\ }%
  \textbf{\bibinfo {volume} {84}},\ \bibinfo {pages} {201103} (\bibinfo {year}
  {2011})%
  \bibAnnoteFile{NoStop}{Ruegg:2011c}%
\bibitem{Haldane:1988}%
  \BibitemOpen
  \bibfield{author}{%
  \bibinfo {author} {\bibfnamefont{F.~D.~M.}\ \bibnamefont{Haldane}},\ }%
  \bibfield{journal}{%
  \Doi{10.1103/PhysRevLett.61.2015}{\bibinfo {journal} {Phys. Rev. Lett.}}\ }%
  \textbf{\bibinfo {volume} {61}},\ \bibinfo {pages} {2015} (\bibinfo {year}
  {1988})%
  \bibAnnoteFile{NoStop}{Haldane:1988}%
\bibitem{Kane:2005b}%
  \BibitemOpen
  \bibfield{author}{%
  \bibinfo {author} {\bibfnamefont{C.~L.}\ \bibnamefont{Kane}}\ and\ \bibinfo
  {author} {\bibfnamefont{E.~J.}\ \bibnamefont{Mele}},\ }%
  \bibfield{journal}{%
  \Doi{10.1103/PhysRevLett.95.146802}{\bibinfo {journal} {Phys. Rev. Lett.}}\
  }%
  \textbf{\bibinfo {volume} {95}},\ \bibinfo {pages} {146802} (\bibinfo {year}
  {2005})%
  \bibAnnoteFile{NoStop}{Kane:2005b}%
\bibitem{Seo:2011}%
  \BibitemOpen
  \bibfield{author}{%
  \bibinfo {author} {\bibfnamefont{H.}~\bibnamefont{Seo}}\ and\ \bibinfo
  {author} {\bibfnamefont{A.~A.}\ \bibnamefont{Demkov}},\ }%
  \bibfield{journal}{%
  \Doi{10.1103/PhysRevB.84.045440}{\bibinfo {journal} {Phys. Rev. B}}\ }%
  \textbf{\bibinfo {volume} {84}},\ \bibinfo {pages} {045440} (\bibinfo {year}
  {2011})%
  \bibAnnoteFile{NoStop}{Seo:2011}%
\bibitem{Hohenberg:1964}%
  \BibitemOpen
  \bibfield{author}{%
  \bibinfo {author} {\bibfnamefont{P.}~\bibnamefont{Hohenberg}}\ and\ \bibinfo
  {author} {\bibfnamefont{W.}~\bibnamefont{Kohn}},\ }%
  \bibfield{journal}{%
  \Doi{10.1103/PhysRev.136.B864}{\bibinfo {journal} {Phys. Rev.}}\ }%
  \textbf{\bibinfo {volume} {136}},\ \bibinfo {pages} {B864} (\bibinfo {month}
  {Nov}\ \bibinfo {year} {1964})%
  \bibAnnoteFile{NoStop}{Hohenberg:1964}%
\bibitem{Kohn:1965}%
  \BibitemOpen
  \bibfield{author}{%
  \bibinfo {author} {\bibfnamefont{W.}~\bibnamefont{Kohn}}\ and\ \bibinfo
  {author} {\bibfnamefont{L.~J.}\ \bibnamefont{Sham}},\ }%
  \bibfield{journal}{%
  \Doi{10.1103/PhysRev.140.A1133}{\bibinfo {journal} {Phys. Rev.}}\ }%
  \textbf{\bibinfo {volume} {140}},\ \bibinfo {pages} {A1133} (\bibinfo {month}
  {Nov}\ \bibinfo {year} {1965})%
  \bibAnnoteFile{NoStop}{Kohn:1965}%
\bibitem{Liu:2011}%
  \BibitemOpen
  \bibfield{author}{%
  \bibinfo {author} {\bibfnamefont{J.}~\bibnamefont{Liu}}, \bibinfo {author}
  {\bibfnamefont{S.}~\bibnamefont{Okamoto}}, \bibinfo {author}
  {\bibfnamefont{M.}~\bibnamefont{van Veenendaal}}, \bibinfo {author}
  {\bibfnamefont{M.}~\bibnamefont{Kareev}}, \bibinfo {author}
  {\bibfnamefont{B.}~\bibnamefont{Gray}}, \bibinfo {author}
  {\bibfnamefont{P.}~\bibnamefont{Ryan}}, \bibinfo {author}
  {\bibfnamefont{J.~W.}\ \bibnamefont{Freeland}},\ and\ \bibinfo {author}
  {\bibfnamefont{J.}~\bibnamefont{Chakhalian}},\ }%
  \bibfield{journal}{%
  \Doi{10.1103/PhysRevB.83.161102}{\bibinfo {journal} {Phys. Rev. B}}\ }%
  \textbf{\bibinfo {volume} {83}},\ \bibinfo {pages} {161102} (\bibinfo {year}
  {2011})%
  \bibAnnoteFile{NoStop}{Liu:2011}%
\bibitem{Boris:2011}%
  \BibitemOpen
  \bibfield{author}{%
  \bibinfo {author} {\bibfnamefont{A.~V.}\ \bibnamefont{Boris}}, \bibinfo
  {author} {\bibfnamefont{Y.}~\bibnamefont{Matiks}}, \bibinfo {author}
  {\bibfnamefont{E.}~\bibnamefont{Benckiser}}, \bibinfo {author}
  {\bibfnamefont{A.}~\bibnamefont{Frano}}, \bibinfo {author}
  {\bibfnamefont{P.}~\bibnamefont{Popovich}}, \bibinfo {author}
  {\bibfnamefont{V.}~\bibnamefont{Hinkov}}, \bibinfo {author}
  {\bibfnamefont{P.}~\bibnamefont{Wochner}}, \bibinfo {author}
  {\bibfnamefont{M.}~\bibnamefont{Castro-Colin}}, \bibinfo {author}
  {\bibfnamefont{E.}~\bibnamefont{Detemple}}, \bibinfo {author}
  {\bibfnamefont{V.~K.}\ \bibnamefont{Malik}}, \bibinfo {author}
  {\bibfnamefont{C.}~\bibnamefont{Bernhard}}, \bibinfo {author}
  {\bibfnamefont{T.}~\bibnamefont{Prokscha}}, \bibinfo {author}
  {\bibfnamefont{A.}~\bibnamefont{Suter}}, \bibinfo {author}
  {\bibfnamefont{Z.}~\bibnamefont{Salman}}, \bibinfo {author}
  {\bibfnamefont{E.}~\bibnamefont{Morenzoni}}, \bibinfo {author}
  {\bibfnamefont{G.}~\bibnamefont{Cristiani}}, \bibinfo {author}
  {\bibfnamefont{H.-U.}\ \bibnamefont{Habermeier}},\ and\ \bibinfo {author}
  {\bibfnamefont{B.}~\bibnamefont{Keimer}},\ }%
  \bibfield{journal}{%
  \Doi{10.1126/science.1202647}{\bibinfo {journal} {Science}}\ }%
  \textbf{\bibinfo {volume} {332}},\ \bibinfo {pages} {937} (\bibinfo {year}
  {2011})%
  \bibAnnoteFile{NoStop}{Boris:2011}%
\bibitem{Freeland:2011}%
  \BibitemOpen
  \bibfield{author}{%
  \bibinfo {author} {\bibfnamefont{J.~W.}\ \bibnamefont{Freeland}}, \bibinfo
  {author} {\bibfnamefont{J.}~\bibnamefont{Liu}}, \bibinfo {author}
  {\bibfnamefont{M.}~\bibnamefont{Kareev}}, \bibinfo {author}
  {\bibfnamefont{B.}~\bibnamefont{Gray}}, \bibinfo {author}
  {\bibfnamefont{J.~W.}\ \bibnamefont{Kim}}, \bibinfo {author}
  {\bibfnamefont{P.}~\bibnamefont{Ryan}}, \bibinfo {author}
  {\bibfnamefont{R.}~\bibnamefont{Pentcheva}},\ and\ \bibinfo {author}
  {\bibfnamefont{J.}~\bibnamefont{Chakhalian}},\ }%
  \bibfield{journal}{%
  \Doi{10.1209/0295-5075/96/57004}{\bibinfo {journal} {EPL (Europhysics
  Letters)}}\ }%
  \textbf{\bibinfo {volume} {96}},\ \bibinfo {pages} {57004} (\bibinfo {year}
  {2011})%
  \bibAnnoteFile{NoStop}{Freeland:2011}%
\bibitem{Chakhalian:2011}%
  \BibitemOpen
  \bibfield{author}{%
  \bibinfo {author} {\bibfnamefont{J.}~\bibnamefont{Chakhalian}}, \bibinfo
  {author} {\bibfnamefont{J.~M.}\ \bibnamefont{Rondinelli}}, \bibinfo {author}
  {\bibfnamefont{J.}~\bibnamefont{Liu}}, \bibinfo {author}
  {\bibfnamefont{B.~A.}\ \bibnamefont{Gray}}, \bibinfo {author}
  {\bibfnamefont{M.}~\bibnamefont{Kareev}}, \bibinfo {author}
  {\bibfnamefont{E.~J.}\ \bibnamefont{Moon}}, \bibinfo {author}
  {\bibfnamefont{N.}~\bibnamefont{Prasai}}, \bibinfo {author}
  {\bibfnamefont{J.~L.}\ \bibnamefont{Cohn}}, \bibinfo {author}
  {\bibfnamefont{M.}~\bibnamefont{Varela}}, \bibinfo {author}
  {\bibfnamefont{I.~C.}\ \bibnamefont{Tung}}, \bibinfo {author}
  {\bibfnamefont{M.~J.}\ \bibnamefont{Bedzyk}}, \bibinfo {author}
  {\bibfnamefont{S.~G.}\ \bibnamefont{Altendorf}}, \bibinfo {author}
  {\bibfnamefont{F.}~\bibnamefont{Strigari}}, \bibinfo {author}
  {\bibfnamefont{B.}~\bibnamefont{Dabrowski}}, \bibinfo {author}
  {\bibfnamefont{L.~H.}\ \bibnamefont{Tjeng}}, \bibinfo {author}
  {\bibfnamefont{P.~J.}\ \bibnamefont{Ryan}},\ and\ \bibinfo {author}
  {\bibfnamefont{J.~W.}\ \bibnamefont{Freeland}},\ }%
  \bibfield{journal}{%
  \Doi{10.1103/PhysRevLett.107.116805}{\bibinfo {journal} {Phys. Rev. Lett.}}\
  }%
  \textbf{\bibinfo {volume} {107}},\ \bibinfo {pages} {116805} (\bibinfo {year}
  {2011})%
  \bibAnnoteFile{NoStop}{Chakhalian:2011}%
\bibitem{Anisimov:1999}%
  \BibitemOpen
  \bibfield{author}{%
  \bibinfo {author} {\bibfnamefont{V.~I.}\ \bibnamefont{Anisimov}}, \bibinfo
  {author} {\bibfnamefont{D.}~\bibnamefont{Bukhvalov}},\ and\ \bibinfo {author}
  {\bibfnamefont{T.~M.}\ \bibnamefont{Rice}},\ }%
  \bibfield{journal}{%
  \Doi{10.1103/PhysRevB.59.7901}{\bibinfo {journal} {Phys. Rev. B}}\ }%
  \textbf{\bibinfo {volume} {59}},\ \bibinfo {pages} {7901} (\bibinfo {year}
  {1999})%
  \bibAnnoteFile{NoStop}{Anisimov:1999}%
\bibitem{Chaloupka:2008}%
  \BibitemOpen
  \bibfield{author}{%
  \bibinfo {author} {\bibfnamefont{J.}~\bibnamefont{Chaloupka}}\ and\ \bibinfo
  {author} {\bibfnamefont{G.}~\bibnamefont{Khaliullin}},\ }%
  \bibfield{journal}{%
  \Doi{10.1103/PhysRevLett.100.016404}{\bibinfo {journal} {Phys. Rev. Lett.}}\
  }%
  \textbf{\bibinfo {volume} {100}},\ \bibinfo {pages} {016404} (\bibinfo {year}
  {2008})%
  \bibAnnoteFile{NoStop}{Chaloupka:2008}%
\bibitem{Hansmann:2009}%
  \BibitemOpen
  \bibfield{author}{%
  \bibinfo {author} {\bibfnamefont{P.}~\bibnamefont{Hansmann}}, \bibinfo
  {author} {\bibfnamefont{X.}~\bibnamefont{Yang}}, \bibinfo {author}
  {\bibfnamefont{A.}~\bibnamefont{Toschi}}, \bibinfo {author}
  {\bibfnamefont{G.}~\bibnamefont{Khaliullin}}, \bibinfo {author}
  {\bibfnamefont{O.~K.}\ \bibnamefont{Andersen}},\ and\ \bibinfo {author}
  {\bibfnamefont{K.}~\bibnamefont{Held}},\ }%
  \bibfield{journal}{%
  \Doi{10.1103/PhysRevLett.103.016401}{\bibinfo {journal} {Phys. Rev. Lett.}}\
  }%
  \textbf{\bibinfo {volume} {103}},\ \bibinfo {pages} {016401} (\bibinfo {year}
  {2009})%
  \bibAnnoteFile{NoStop}{Hansmann:2009}%
\bibitem{Zhou:2003}%
  \BibitemOpen
  \bibfield{author}{%
  \bibinfo {author} {\bibfnamefont{J.-S.}\ \bibnamefont{Zhou}}, \bibinfo
  {author} {\bibfnamefont{J.~B.}\ \bibnamefont{Goodenough}},\ and\ \bibinfo
  {author} {\bibfnamefont{B.}~\bibnamefont{Dabrowski}},\ }%
  \bibfield{journal}{%
  \Doi{10.1103/PhysRevB.67.020404}{\bibinfo {journal} {Phys. Rev. B}}\ }%
  \textbf{\bibinfo {volume} {67}},\ \bibinfo {pages} {020404} (\bibinfo {year}
  {2003})%
  \bibAnnoteFile{NoStop}{Zhou:2003}%
\bibitem{Sun:2009}%
  \BibitemOpen
  \bibfield{author}{%
  \bibinfo {author} {\bibfnamefont{K.}~\bibnamefont{Sun}}, \bibinfo {author}
  {\bibfnamefont{H.}~\bibnamefont{Yao}}, \bibinfo {author}
  {\bibfnamefont{E.}~\bibnamefont{Fradkin}},\ and\ \bibinfo {author}
  {\bibfnamefont{S.~A.}\ \bibnamefont{Kivelson}},\ }%
  \bibfield{journal}{%
  \Doi{10.1103/PhysRevLett.103.046811}{\bibinfo {journal} {Phys. Rev. Lett.}}\
  }%
  \textbf{\bibinfo {volume} {103}},\ \bibinfo {eid} {046811} (\bibinfo {year}
  {2009})%
  \bibAnnoteFile{NoStop}{Sun:2009}%
\bibitem{Dudarev:1998}%
  \BibitemOpen
  \bibfield{author}{%
  \bibinfo {author} {\bibfnamefont{S.~L.}\ \bibnamefont{Dudarev}}, \bibinfo
  {author} {\bibfnamefont{G.~A.}\ \bibnamefont{Botton}}, \bibinfo {author}
  {\bibfnamefont{S.~Y.}\ \bibnamefont{Savrasov}}, \bibinfo {author}
  {\bibfnamefont{C.~J.}\ \bibnamefont{Humphreys}},\ and\ \bibinfo {author}
  {\bibfnamefont{A.~P.}\ \bibnamefont{Sutton}},\ }%
  \bibfield{journal}{%
  \Doi{10.1103/PhysRevB.57.1505}{\bibinfo {journal} {Phys. Rev. B}}\ }%
  \textbf{\bibinfo {volume} {57}},\ \bibinfo {pages} {1505} (\bibinfo {year}
  {1998})%
  \bibAnnoteFile{NoStop}{Dudarev:1998}%
\bibitem{Zaanen:1985}%
  \BibitemOpen
  \bibfield{author}{%
  \bibinfo {author} {\bibfnamefont{J.}~\bibnamefont{Zaanen}}, \bibinfo {author}
  {\bibfnamefont{G.~A.}\ \bibnamefont{Sawatzky}},\ and\ \bibinfo {author}
  {\bibfnamefont{J.~W.}\ \bibnamefont{Allen}},\ }%
  \bibfield{journal}{%
  \Doi{10.1103/PhysRevLett.55.418}{\bibinfo {journal} {Phys. Rev. Lett.}}\ }%
  \textbf{\bibinfo {volume} {55}},\ \bibinfo {pages} {418} (\bibinfo {year}
  {1985})%
  \bibAnnoteFile{NoStop}{Zaanen:1985}%
\bibitem{Barth:1972}%
  \BibitemOpen
  \bibfield{author}{%
  \bibinfo {author} {\bibfnamefont{U.}~\bibnamefont{von Barth}}\ and\ \bibinfo
  {author} {\bibfnamefont{L.}~\bibnamefont{Hedin}},\ }%
  \bibfield{journal}{%
  \Doi{10.1088/0022-3719/5/13/012}{\bibinfo {journal} {Journal of Physics C:
  Solid State Physics}}\ }%
  \textbf{\bibinfo {volume} {5}},\ \bibinfo {pages} {1629} (\bibinfo {year}
  {1972})%
  \bibAnnoteFile{NoStop}{Barth:1972}%
\bibitem{Kresse:1996}%
  \BibitemOpen
  \bibfield{author}{%
  \bibinfo {author} {\bibfnamefont{G.}~\bibnamefont{Kresse}}\ and\ \bibinfo
  {author} {\bibfnamefont{J.}~\bibnamefont{Furthm\"uller}},\ }%
  \bibfield{journal}{%
  \Doi{10.1103/PhysRevB.54.11169}{\bibinfo {journal} {Phys. Rev. B}}\ }%
  \textbf{\bibinfo {volume} {54}},\ \bibinfo {pages} {11169} (\bibinfo {year}
  {1996})%
  \bibAnnoteFile{NoStop}{Kresse:1996}%
\bibitem{Blochl:1994}%
  \BibitemOpen
  \bibfield{author}{%
  \bibinfo {author} {\bibfnamefont{P.~E.}\ \bibnamefont{Bl\"ochl}},\ }%
  \bibfield{journal}{%
  \Doi{10.1103/PhysRevB.50.17953}{\bibinfo {journal} {Phys. Rev. B}}\ }%
  \textbf{\bibinfo {volume} {50}},\ \bibinfo {pages} {17953} (\bibinfo {year}
  {1994})%
  \bibAnnoteFile{NoStop}{Blochl:1994}%
\bibitem{Gou:2011}%
  \BibitemOpen
  \bibfield{author}{%
  \bibinfo {author} {\bibfnamefont{G.}~\bibnamefont{Gou}}, \bibinfo {author}
  {\bibfnamefont{I.}~\bibnamefont{Grinberg}}, \bibinfo {author}
  {\bibfnamefont{A.~M.}\ \bibnamefont{Rappe}},\ and\ \bibinfo {author}
  {\bibfnamefont{J.~M.}\ \bibnamefont{Rondinelli}},\ }%
  \bibfield{journal}{%
  \Doi{10.1103/PhysRevB.84.144101}{\bibinfo {journal} {Phys. Rev. B}}\ }%
  \textbf{\bibinfo {volume} {84}},\ \bibinfo {pages} {144101} (\bibinfo {year}
  {2011})%
  \bibAnnoteFile{NoStop}{Gou:2011}%
\bibitem{Cococcioni:2005}%
  \BibitemOpen
  \bibfield{author}{%
  \bibinfo {author} {\bibfnamefont{M.}~\bibnamefont{Cococcioni}}\ and\ \bibinfo
  {author} {\bibfnamefont{S.}~\bibnamefont{de~Gironcoli}},\ }%
  \bibfield{journal}{%
  \Doi{10.1103/PhysRevB.71.035105}{\bibinfo {journal} {Phys. Rev. B}}\ }%
  \textbf{\bibinfo {volume} {71}},\ \bibinfo {pages} {035105} (\bibinfo {year}
  {2005})%
  \bibAnnoteFile{NoStop}{Cococcioni:2005}%
\bibitem{Nohara:2009}%
  \BibitemOpen
  \bibfield{author}{%
  \bibinfo {author} {\bibfnamefont{Y.}~\bibnamefont{Nohara}}, \bibinfo {author}
  {\bibfnamefont{S.}~\bibnamefont{Yamamoto}},\ and\ \bibinfo {author}
  {\bibfnamefont{T.}~\bibnamefont{Fujiwara}},\ }%
  \bibfield{journal}{%
  \Doi{10.1103/PhysRevB.79.195110}{\bibinfo {journal} {Phys. Rev. B}}\ }%
  \textbf{\bibinfo {volume} {79}},\ \bibinfo {pages} {195110} (\bibinfo {year}
  {2009})%
  \bibAnnoteFile{NoStop}{Nohara:2009}%
\bibitem{Abbate:2002}%
  \BibitemOpen
  \bibfield{author}{%
  \bibinfo {author} {\bibfnamefont{M.}~\bibnamefont{Abbate}}, \bibinfo {author}
  {\bibfnamefont{G.}~\bibnamefont{Zampieri}}, \bibinfo {author}
  {\bibfnamefont{F.}~\bibnamefont{Prado}}, \bibinfo {author}
  {\bibfnamefont{A.}~\bibnamefont{Caneiro}}, \bibinfo {author}
  {\bibfnamefont{J.~M.}\ \bibnamefont{Gonzalez-Calbet}},\ and\ \bibinfo
  {author} {\bibfnamefont{M.}~\bibnamefont{Vallet-Regi}},\ }%
  \bibfield{journal}{%
  \Doi{10.1103/PhysRevB.65.155101}{\bibinfo {journal} {Phys. Rev. B}}\ }%
  \textbf{\bibinfo {volume} {65}},\ \bibinfo {pages} {155101} (\bibinfo {year}
  {2002})%
  \bibAnnoteFile{NoStop}{Abbate:2002}%
\bibitem{Slater:1954}%
  \BibitemOpen
  \bibfield{author}{%
  \bibinfo {author} {\bibfnamefont{J.~C.}\ \bibnamefont{Slater}}\ and\ \bibinfo
  {author} {\bibfnamefont{G.~F.}\ \bibnamefont{Koster}},\ }%
  \bibfield{journal}{%
  \Doi{10.1103/PhysRev.94.1498}{\bibinfo {journal} {Phys. Rev.}}\ }%
  \textbf{\bibinfo {volume} {94}},\ \bibinfo {pages} {1498} (\bibinfo {year}
  {1954})%
  \bibAnnoteFile{NoStop}{Slater:1954}%
\bibitem{Wu:2007}%
  \BibitemOpen
  \bibfield{author}{%
  \bibinfo {author} {\bibfnamefont{C.}~\bibnamefont{Wu}}, \bibinfo {author}
  {\bibfnamefont{D.}~\bibnamefont{Bergman}}, \bibinfo {author}
  {\bibfnamefont{L.}~\bibnamefont{Balents}},\ and\ \bibinfo {author}
  {\bibfnamefont{S.}~\bibnamefont{Das~Sarma}},\ }%
  \bibfield{journal}{%
  \Doi{10.1103/PhysRevLett.99.070401}{\bibinfo {journal} {Phys. Rev. Lett.}}\
  }%
  \textbf{\bibinfo {volume} {99}},\ \bibinfo {pages} {070401} (\bibinfo {year}
  {2007})%
  \bibAnnoteFile{NoStop}{Wu:2007}%
\bibitem{Note1}%
  \BibitemOpen
  \bibinfo {note} {We would like to thank Adrian Feiguin for pointing out a
  possible connection.}%
  \bibAnnoteFile{Stop}{Note1}%
\bibitem{Bergman:2008}%
  \BibitemOpen
  \bibfield{author}{%
  \bibinfo {author} {\bibfnamefont{D.~L.}\ \bibnamefont{Bergman}}, \bibinfo
  {author} {\bibfnamefont{C.}~\bibnamefont{Wu}},\ and\ \bibinfo {author}
  {\bibfnamefont{L.}~\bibnamefont{Balents}},\ }%
  \bibfield{journal}{%
  \Doi{10.1103/PhysRevB.78.125104}{\bibinfo {journal} {Phys. Rev. B}}\ }%
  \textbf{\bibinfo {volume} {78}},\ \bibinfo {pages} {125104} (\bibinfo {year}
  {2008})%
  \bibAnnoteFile{NoStop}{Bergman:2008}%
\bibitem{Wen:2010}%
  \BibitemOpen
  \bibfield{author}{%
  \bibinfo {author} {\bibfnamefont{J.}~\bibnamefont{Wen}}, \bibinfo {author}
  {\bibfnamefont{A.}~\bibnamefont{R\"uegg}}, \bibinfo {author}
  {\bibfnamefont{C.~C.~J.}\ \bibnamefont{Wang}},\ and\ \bibinfo {author}
  {\bibfnamefont{G.~A.}\ \bibnamefont{Fiete}},\ }%
  \bibfield{journal}{%
  \Doi{10.1103/PhysRevB.82.075125}{\bibinfo {journal} {Phys. Rev. B}}\ }%
  \textbf{\bibinfo {volume} {82}},\ \bibinfo {pages} {075125} (\bibinfo {year}
  {2010})%
  \bibAnnoteFile{NoStop}{Wen:2010}%
\bibitem{Uebelacker:2011}%
  \BibitemOpen
  \bibfield{author}{%
  \bibinfo {author} {\bibfnamefont{S.}~\bibnamefont{Uebelacker}}\ and\ \bibinfo
  {author} {\bibfnamefont{C.}~\bibnamefont{Honerkamp}},\ }%
  \bibfield{journal}{%
  \Doi{10.1103/PhysRevB.84.205122}{\bibinfo {journal} {Phys. Rev. B}}\ }%
  \textbf{\bibinfo {volume} {84}},\ \bibinfo {pages} {205122} (\bibinfo {year}
  {2011})%
  \bibAnnoteFile{NoStop}{Uebelacker:2011}%
\bibitem{LeeS:2011b}%
  \BibitemOpen
  \bibfield{author}{%
  \bibinfo {author} {\bibfnamefont{S.~B.}\ \bibnamefont{Lee}}, \bibinfo
  {author} {\bibfnamefont{R.}~\bibnamefont{Chen}},\ and\ \bibinfo {author}
  {\bibfnamefont{L.}~\bibnamefont{Balents}},\ }%
  \bibfield{journal}{%
  \Doi{10.1103/PhysRevB.84.165119}{\bibinfo {journal} {Phys. Rev. B}}\ }%
  \textbf{\bibinfo {volume} {84}},\ \bibinfo {pages} {165119} (\bibinfo {month}
  {Oct}\ \bibinfo {year} {2011})%
  \bibAnnoteFile{NoStop}{LeeS:2011b}%
\bibitem{Mizokawa:1996}%
  \BibitemOpen
  \bibfield{author}{%
  \bibinfo {author} {\bibfnamefont{T.}~\bibnamefont{Mizokawa}}\ and\ \bibinfo
  {author} {\bibfnamefont{A.}~\bibnamefont{Fujimori}},\ }%
  \bibfield{journal}{%
  \Doi{10.1103/PhysRevB.54.5368}{\bibinfo {journal} {Phys. Rev. B}}\ }%
  \textbf{\bibinfo {volume} {54}},\ \bibinfo {pages} {5368} (\bibinfo {year}
  {1996})%
  \bibAnnoteFile{NoStop}{Mizokawa:1996}%
\bibitem{Imada:1998}%
  \BibitemOpen
  \bibfield{author}{%
  \bibinfo {author} {\bibfnamefont{M.}~\bibnamefont{Imada}}, \bibinfo {author}
  {\bibfnamefont{A.}~\bibnamefont{Fujimori}},\ and\ \bibinfo {author}
  {\bibfnamefont{Y.}~\bibnamefont{Tokura}},\ }%
  \bibfield{journal}{%
  \Doi{10.1103/RevModPhys.70.1039}{\bibinfo {journal} {Rev. Mod. Phys.}}\ }%
  \textbf{\bibinfo {volume} {70}},\ \bibinfo {pages} {1039} (\bibinfo {year}
  {1998})%
  \bibAnnoteFile{NoStop}{Imada:1998}%
\bibitem{Nasu:2008}%
  \BibitemOpen
  \bibfield{author}{%
  \bibinfo {author} {\bibfnamefont{J.}~\bibnamefont{Nasu}}, \bibinfo {author}
  {\bibfnamefont{A.}~\bibnamefont{Nagano}}, \bibinfo {author}
  {\bibfnamefont{M.}~\bibnamefont{Naka}},\ and\ \bibinfo {author}
  {\bibfnamefont{S.}~\bibnamefont{Ishihara}},\ }%
  \bibfield{journal}{%
  \Doi{10.1103/PhysRevB.78.024416}{\bibinfo {journal} {Phys. Rev. B}}\ }%
  \textbf{\bibinfo {volume} {78}},\ \bibinfo {pages} {024416} (\bibinfo {year}
  {2008})%
  \bibAnnoteFile{NoStop}{Nasu:2008}%
\bibitem{Moessner:2001}%
  \BibitemOpen
  \bibfield{author}{%
  \bibinfo {author} {\bibfnamefont{R.}~\bibnamefont{Moessner}}, \bibinfo
  {author} {\bibfnamefont{S.~L.}\ \bibnamefont{Sondhi}},\ and\ \bibinfo
  {author} {\bibfnamefont{P.}~\bibnamefont{Chandra}},\ }%
  \bibfield{journal}{%
  \Doi{10.1103/PhysRevB.64.144416}{\bibinfo {journal} {Phys. Rev. B}}\ }%
  \textbf{\bibinfo {volume} {64}},\ \bibinfo {pages} {144416} (\bibinfo {year}
  {2001})%
  \bibAnnoteFile{NoStop}{Moessner:2001}%
\bibitem{OBrien:2010}%
  \BibitemOpen
  \bibfield{author}{%
  \bibinfo {author} {\bibfnamefont{A.}~\bibnamefont{O'Brien}}, \bibinfo
  {author} {\bibfnamefont{F.}~\bibnamefont{Pollmann}},\ and\ \bibinfo {author}
  {\bibfnamefont{P.}~\bibnamefont{Fulde}},\ }%
  \bibfield{journal}{%
  \Doi{10.1103/PhysRevB.81.235115}{\bibinfo {journal} {Phys. Rev. B}}\ }%
  \textbf{\bibinfo {volume} {81}},\ \bibinfo {pages} {235115} (\bibinfo {year}
  {2010})%
  \bibAnnoteFile{NoStop}{OBrien:2010}%
\bibitem{Ruegg:2011}%
  \BibitemOpen
  \bibfield{author}{%
  \bibinfo {author} {\bibfnamefont{A.}~\bibnamefont{R\"uegg}}\ and\ \bibinfo
  {author} {\bibfnamefont{G.~A.}\ \bibnamefont{Fiete}},\ }%
  \bibfield{journal}{%
  \Doi{10.1103/PhysRevB.83.165118}{\bibinfo {journal} {Phys. Rev. B}}\ }%
  \textbf{\bibinfo {volume} {83}},\ \bibinfo {pages} {165118} (\bibinfo {year}
  {2011})%
  \bibAnnoteFile{NoStop}{Ruegg:2011}%
\bibitem{Pollmann:2006}%
  \BibitemOpen
  \bibfield{author}{%
  \bibinfo {author} {\bibfnamefont{F.}~\bibnamefont{Pollmann}}, \bibinfo
  {author} {\bibfnamefont{P.}~\bibnamefont{Fulde}},\ and\ \bibinfo {author}
  {\bibfnamefont{E.}~\bibnamefont{Runge}},\ }%
  \bibfield{journal}{%
  \Doi{10.1103/PhysRevB.73.125121}{\bibinfo {journal} {Phys. Rev. B}}\ }%
  \textbf{\bibinfo {volume} {73}},\ \bibinfo {pages} {125121} (\bibinfo {year}
  {2006})%
  \bibAnnoteFile{NoStop}{Pollmann:2006}%
\bibitem{Thouless:1982}%
  \BibitemOpen
  \bibfield{author}{%
  \bibinfo {author} {\bibfnamefont{D.~J.}\ \bibnamefont{Thouless}}, \bibinfo
  {author} {\bibfnamefont{M.}~\bibnamefont{Kohmoto}}, \bibinfo {author}
  {\bibfnamefont{M.~P.}\ \bibnamefont{Nightingale}},\ and\ \bibinfo {author}
  {\bibfnamefont{M.}~\bibnamefont{den Nijs}},\ }%
  \bibfield{journal}{%
  \Doi{10.1103/PhysRevLett.49.405}{\bibinfo {journal} {Phys. Rev. Lett.}}\ }%
  \textbf{\bibinfo {volume} {49}},\ \bibinfo {pages} {405} (\bibinfo {year}
  {1982})%
  \bibAnnoteFile{NoStop}{Thouless:1982}%
\bibitem{Mazin:2007}%
  \BibitemOpen
  \bibfield{author}{%
  \bibinfo {author} {\bibfnamefont{I.~I.}\ \bibnamefont{Mazin}}, \bibinfo
  {author} {\bibfnamefont{D.~I.}\ \bibnamefont{Khomskii}}, \bibinfo {author}
  {\bibfnamefont{R.}~\bibnamefont{Lengsdorf}}, \bibinfo {author}
  {\bibfnamefont{J.~A.}\ \bibnamefont{Alonso}}, \bibinfo {author}
  {\bibfnamefont{W.~G.}\ \bibnamefont{Marshall}}, \bibinfo {author}
  {\bibfnamefont{R.~M.}\ \bibnamefont{Ibberson}}, \bibinfo {author}
  {\bibfnamefont{A.}~\bibnamefont{Podlesnyak}}, \bibinfo {author}
  {\bibfnamefont{M.~J.}\ \bibnamefont{Martinez-Lope}},\ and\ \bibinfo {author}
  {\bibfnamefont{M.~M.}\ \bibnamefont{Abd-Elmeguid}},\ }%
  \bibfield{journal}{%
  \Doi{10.1103/PhysRevLett.98.176406}{\bibinfo {journal} {Phys. Rev. Lett.}}\
  }%
  \textbf{\bibinfo {volume} {98}},\ \bibinfo {pages} {176406} (\bibinfo {year}
  {2007})%
  \bibAnnoteFile{NoStop}{Mazin:2007}%
\bibitem{Han:2010}%
  \BibitemOpen
  \bibfield{author}{%
  \bibinfo {author} {\bibfnamefont{M.~J.}\ \bibnamefont{Han}}, \bibinfo
  {author} {\bibfnamefont{C.~A.}\ \bibnamefont{Marianetti}},\ and\ \bibinfo
  {author} {\bibfnamefont{A.~J.}\ \bibnamefont{Millis}},\ }%
  \bibfield{journal}{%
  \Doi{10.1103/PhysRevB.82.134408}{\bibinfo {journal} {Phys. Rev. B}}\ }%
  \textbf{\bibinfo {volume} {82}},\ \bibinfo {pages} {134408} (\bibinfo {year}
  {2010})%
  \bibAnnoteFile{NoStop}{Han:2010}%
\bibitem{Anisimov:1997}%
  \BibitemOpen
  \bibfield{author}{%
  \bibinfo {author} {\bibfnamefont{V.~I.}\ \bibnamefont{Anisimov}}, \bibinfo
  {author} {\bibfnamefont{F.}~\bibnamefont{Aryasetiawan}},\ and\ \bibinfo
  {author} {\bibfnamefont{A.~I.}\ \bibnamefont{Lichtenstein}},\ }%
  \bibfield{journal}{%
  \Doi{10.1088/0953-8984/9/4/002}{\bibinfo {journal} {Journal of Physics:
  Condensed Matter}}\ }%
  \textbf{\bibinfo {volume} {9}},\ \bibinfo {pages} {767} (\bibinfo {year}
  {1997})%
  \bibAnnoteFile{NoStop}{Anisimov:1997}%
\bibitem{Kotliar:2006}%
  \BibitemOpen
  \bibfield{author}{%
  \bibinfo {author} {\bibfnamefont{G.}~\bibnamefont{Kotliar}}, \bibinfo
  {author} {\bibfnamefont{S.~Y.}\ \bibnamefont{Savrasov}}, \bibinfo {author}
  {\bibfnamefont{K.}~\bibnamefont{Haule}}, \bibinfo {author}
  {\bibfnamefont{V.~S.}\ \bibnamefont{Oudovenko}}, \bibinfo {author}
  {\bibfnamefont{O.}~\bibnamefont{Parcollet}},\ and\ \bibinfo {author}
  {\bibfnamefont{C.~A.}\ \bibnamefont{Marianetti}},\ }%
  \bibfield{journal}{%
  \Doi{10.1103/RevModPhys.78.865}{\bibinfo {journal} {Rev. Mod. Phys.}}\ }%
  \textbf{\bibinfo {volume} {78}},\ \bibinfo {eid} {865} (\bibinfo {year}
  {2006})%
  \bibAnnoteFile{NoStop}{Kotliar:2006}%
\bibitem{Mizokawa:2000}%
  \BibitemOpen
  \bibfield{author}{%
  \bibinfo {author} {\bibfnamefont{T.}~\bibnamefont{Mizokawa}}, \bibinfo
  {author} {\bibfnamefont{D.~I.}\ \bibnamefont{Khomskii}},\ and\ \bibinfo
  {author} {\bibfnamefont{G.~A.}\ \bibnamefont{Sawatzky}},\ }%
  \bibfield{journal}{%
  \Doi{10.1103/PhysRevB.61.11263}{\bibinfo {journal} {Phys. Rev. B}}\ }%
  \textbf{\bibinfo {volume} {61}},\ \bibinfo {pages} {11263} (\bibinfo {year}
  {2000})%
  \bibAnnoteFile{NoStop}{Mizokawa:2000}%
\bibitem{Khomskii:2001}%
  \BibitemOpen
  \bibfield{author}{%
  \bibinfo {author} {\bibfnamefont{D.}~\bibnamefont{{Khomskii}}},\ }%
  \bibfield{journal}{%
  \bibinfo {journal} {ArXiv eprints}}%
   (\bibinfo {month} {Jan.}\ \bibinfo {year} {2001}),\
  \Eprint{http://arxiv.org/abs/arXiv:cond-mat/0101164}{arXiv:cond-mat/0101164}%
  \bibAnnoteFile{NoStop}{Khomskii:2001}%
\bibitem{Wangb:2011}%
  \BibitemOpen
  \bibfield{author}{%
  \bibinfo {author} {\bibfnamefont{X.}~\bibnamefont{{Wang}}}, \bibinfo {author}
  {\bibfnamefont{M.~J.}\ \bibnamefont{{Han}}}, \bibinfo {author}
  {\bibfnamefont{L.}~\bibnamefont{{de' Medici}}}, \bibinfo {author}
  {\bibfnamefont{C.~A.}\ \bibnamefont{{Marianetti}}},\ and\ \bibinfo {author}
  {\bibfnamefont{A.~J.}\ \bibnamefont{{Millis}}},\ }%
  \bibfield{journal}{%
  \bibinfo {journal} {ArXiv e-prints}}%
   (\bibinfo {month} {Oct.}\ \bibinfo {year} {2011}),\
  \Eprint{http://arxiv.org/abs/1110.2782}{arXiv:1110.2782}%
  \bibAnnoteFile{NoStop}{Wangb:2011}%
\bibitem{Han:2011}%
  \BibitemOpen
  \bibfield{author}{%
  \bibinfo {author} {\bibfnamefont{M.~J.}\ \bibnamefont{Han}}, \bibinfo
  {author} {\bibfnamefont{X.}~\bibnamefont{Wang}}, \bibinfo {author}
  {\bibfnamefont{C.~A.}\ \bibnamefont{Marianetti}},\ and\ \bibinfo {author}
  {\bibfnamefont{A.~J.}\ \bibnamefont{Millis}},\ }%
  \bibfield{journal}{%
  \Doi{10.1103/PhysRevLett.107.206804}{\bibinfo {journal} {Phys. Rev. Lett.}}\
  }%
  \textbf{\bibinfo {volume} {107}},\ \bibinfo {pages} {206804} (\bibinfo {year}
  {2011})%
  \bibAnnoteFile{NoStop}{Han:2011}%
\end{thebibliography}%

\end{document}